\documentstyle[eqsecnum,aps]{revtex}

\input{epsf.tex}
\begin{document}
\draft
\twocolumn[\hsize\textwidth\columnwidth\hsize\csname
@twocolumnfalse\endcsname
\preprint{HEP/123-qed}
\renewcommand{\thefootnote}{\alph{footnote}} 
\title{Phase Mixing in Unperturbed and Perturbed Hamiltonian Systems}
\author{Henry E. Kandrup\footnote{Electronic address: kandrup@astro.ufl.edu}}
\address{ Department of Astronomy, Department of Physics, and
Institute for Fundamental Theory
\\
University of Florida, Gainesville, Florida 32611}
\author{Steven J. Novotny\footnote{Electronic address: sideris@astro.ufl.edu}}
\address{Department of Astronomy, University of Florida, Gainesville, 
Florida 32611\\}

\date{\today}
\maketitle
\begin{abstract}
This paper summarises a numerical investigation of phase mixing in 
time-independent Hamiltonian systems that admit a coexistence of regular
and chaotic phase space regions, allowing also for low
amplitude perturbations idealised as periodic driving, friction, and/or
white and colored noise. The evolution of initially localised ensembles
of orbits was probed through lower order moments and coarse-grained
distribution functions. In the absence of time-dependent perturbations,
regular ensembles disperse initially as a power law in time and only 
exhibit a coarse-grained approach towards an invariant equilibrium over
comparatively long times. Chaotic ensembles generally diverge exponentially 
fast on a time scale related to a typical finite time Lyapunov exponent, but 
can exhibit complex behaviour if they are impacted by the effects of cantori
or the Arnold web. Viewed over somewhat longer times, chaotic ensembles 
typical converge exponentially towards an invariant or near-invariant 
equilibrium. This, however, need not correspond to a {\em true} equilibrium, 
which may only be approached over {\em very} long time scales. Time-dependent 
perturbations can dramatically increase the efficiency of phase mixing, both 
by accelerating the approach towards a near-equilibrium and
by facilitating diffusion through cantori or along the Arnold web 
so as to accelerate the approach towards a true equilibrium. The efficacy of 
such perturbations typically scales logarithmically in amplitude, but is 
comparatively insensitive to most other details, a conclusion which reinforces 
the interpretation that the perturbations act via a resonant coupling.
\end{abstract}
]
\narrowtext
\section{INTRODUCTION AND MOTIVATION}
 \label{sec:level1}
Phase mixing is a fundamental process acting in many-particle systems, with 
important implications for a variety of different areas of physics extending
from galactic dynamics\cite{Ber} to the physics of charged particle 
beams\cite{Rei}. In response to external forces and/or self-consistent
interactions, initially localised collections of particles often tend to 
disperse. Phase mixing is believed to play a fundamental role in so-called 
``violent relaxation,'' the process whereby the stars in an elliptical galaxy 
evolve towards a statistical meta-equilibrium and, under certain 
circumstances, it can play a major role in the dissipation of accelerator 
beams, resulting in substantial -- and undesireable -- increases in emittance. 

Until recently, most practical considerations of phase mixing have assumed, at
least implicitly, that the global dynamics is integrable 
or near-integrable, in which case initially localised ensembles typically 
disperse as a power law in time. However, there is emerging evidence that, in
at least some settings relevant to galactic astronomy\cite{MF} and accelerator
dynamics\cite{Kis}, one is confronted with strongly nonintegrable chaotic 
dynamics. The manifestations of phase mixing in chaotic systems are very 
different from those in nearly integrable systems\cite{KM}, a fact which has
led galactic astronomers to speak of a new phenomenon called {\em chaotic
mixing}\cite{MV}. It is, {\em e.g.,} obvious that localised ensembles of 
chaotic orbits will typically disperse exponentially rather than as a power 
law in time. For this reason, it would seem important to re-examine  
the physics of phase mixing, allowing explicitly for systems
that admits large measures of chaos. 

Of especial interest are the potential implications of chaotic phase space 
mixing for relaxation towards a statistical equilibrium or near-equilibrium.
By reformulating the dynamics of a time-independent Hamiltonian system as
a geodesic flow, {\em e.g.,} using Maupertuis' Principle\cite{Arn}, some 
rigorous mathematical results have been derived. For example, it is 
known that initially localised ensembles of geodesics on a compact manifold 
with constant negative curvature exhibit a coarse-grained approach, 
exponential in time, towards a uniform population of the manifold, 
{\em i.e.,} a microcanonical distribution, at a rate ${\lambda}$ that is 
fixed completely by the magnitude of the curvature\cite{Hopf}. This implies 
the existence of a direct connection between ${\lambda}$ and the positive
Lyapunov exponent(s) ${\chi}$. If the curvature is always negative but not 
constant, the flow becomes significantly more complex, but one can still infer 
an exponential approach towards a microcanonical equilibrium at a rate that is 
bounded from below by the least negative value of the curvature\cite{Ano}. 

However, these results are of limited applicability to real physical systems 
where the curvature can vary in sign and, in many cases, may be 
strictly positive\cite{pos}. In such systems, it would appear\cite{Pet} that 
chaos should be attributed to parametric instability rather than to negative 
curvature, the division of a constant energy hypersurface into chaotic and
regular regions reflecting properties of the local curvature which do, or 
do not, trigger an instability. Within a suitably 
defined connected chaotic phase space region, ensembles may still relax 
exponentially towards a microcanonical population\cite{KM}, but there is no 
reason to expect a one-to-one connection between the rate ${\lambda}$ and the 
Lyapunov exponents ${\chi}$.

Additional complications can also arise
when the bulk dynamics is impacted significantly by phase space 
structures like cantori or an Arnold web\cite{LL1}. If a connected chaotic 
phase space region is partitioned by such {\em entropy barriers}, {\em i.e.,}
phase space obstructions which impede, but do not completely block, orbital
motions, phase space transport can entail a multi-stage process characterised
by two or more vastly different time scales. In analogy with the classical
problem of effusion of gas through a tiny hole, particles starting on one side
of a barrier can evolve towards a near-uniform population of the phase space
region on that side on a time scale $t_{s}$ much shorter than the time scale
$t_{l}$ on which particles breech the barrier. On intermediate time scales,
$t_{s}{\;}{\ll}{\;}t{\;}{\ll}{\;}t_{l}$, each side may be in a near-equilibrium
but the densities on the two sides will in general be very different.

Significantly, however, even very small 
perturbations can accelerate phase space transport through such barriers.
It has been long known from the study of maps\cite{LL2}
that small random perturbations, {\em i.e.,} noise, can dramatically reduce
the time scale for diffusion through cantori. And similarly, it has been 
recognised that even very low amplitude periodic driving can have important 
implications for phase space transport in systems like plasmas\cite{Ten}. 
Recent work on continuous Hamiltonian systems\cite{PK,SK} has served 
to quantify this effect, demonstrating that weak noise and/or periodic driving 
can significantly impact phase space transport on time scales much shorter than
the relaxation time $t_{R}$ on which they significantly change the
values of quantities like energy, which would be conserved absolutely in the
absence of perturbations. 
All this would suggest that low amplitude perturbations can significantly
impact the phase mixing of chaotic orbits, a possibility that has direct
physical implications for the destabilisation of various structures in 
a galaxy\cite{PAQ} or the defocusing of an accelerator beam\cite{Hab}.

This paper presents the results of a systematic numerical
investigation of phase mixing for two- and three-degrees-of-freedom 
Hamiltonian systems that admit a complex coexistence of regular and
chaotic phase space regions, allowing also for the effects of low amplitude
perturbations idealised as periodic driving and/or friction and (in general
coloured) noise. 

In principle, one might argue that such an analysis is not
completely applicable to real many-body systems where the dimensionality of
the true phase space is much larger. 
However, there are at least three reasons to believe 
that the analysis summarised in this paper should find direct physical 
applications. 

1. In a variety of physical systems, {\em e.g.,} relatively low intensity 
beams\cite{Rei} or stars in the center of a galaxy orbiting close to a
supermassive black hole, one expects that the bulk properties of the flow will 
be dominated by `external' bulk forces rather than self-interactions; and, to 
the extent that self-interactions {\em are} important, one might hope to model 
their effects as friction and noise in the context of a Fokker-Planck 
description.

2. Recent integrations of gravitationally interacting test 
particles in fixed, time-independent $N$-body realisations of smooth density 
distributions have shown that, at least for large $N$, these particles 
behave in a fashion closely resembling the motions of particles evolved 
in the corresponding smooth density distribution; and that the differences
that {\em do} exist decrease with increasing $N$\cite{VM,KS2,KS3}. Moreover,
in terms of both the statistical properties of individual orbits {\em and}
the phase mixing of orbit ensembles, discreteness effects are 
extremely well mimicked by (a suitably defined) Gaussian white noise\cite{KS3}.

3. Performing honest $N$-body integrations of interacting systems for very
large $N$ is not viable using current computational resources. However,
fully self-consistent grid code simulations of intense charged beams 
have been found to exhibit both regular and 
chaotic phase mixing qualitatively similar to what has been found in smooth
two- and three-dimensional potentials\cite{Kis,Bohn}.

White noise appears to be very successful in modelling discreteness effects 
in $N$-body systems, where the time scale on which the `random' 
forces change is extremely short. However, real systems are also subjected to 
other, more slowly varying, perturbations which, in many cases, one might hope 
to model as (i) periodic driving and/or (ii) coloured noise with a finite 
autocorrelation time, {\em i.e.,} `random' forces that only change 
significantly on macroscopic time scales.

In the context of galactic astronomy, periodic driving could reflect the effect
of companion galaxies, such as the Small and Large Magellanic Clouds orbiting 
the Milky Way. Viewed mathematically, coloured noise is a superposition of 
periodic perturbations combined with random phases. One might, therefore, use 
coloured noise to model a high density cluster environment, in which, over the 
course of time, the perturbations acting on any given galaxy result from 
complex interactions with a variety of different neighbouring galaxies. 
Alternatively, coloured noise might be used to model discrete substructures 
like molecular clouds or globular clusters, where the interaction time scale 
is too long to justify a white noise approximation. 

Such perturbations may also influence the dynamics of charged-particle beams.
In a cyclotron or synchrotron, irregularities in the confining and/or
focusing magnetic fields would, from the perspective of the beam, appear as
quasi-periodic perturbations with, perhaps, a slowly varying frequency. The 
conceptual paradigm for such beams is the H\'enon-Heiles potential,
 $V(x,y)={1\over 2}(x^{2}+y^{2}+2x^{2}y-{2\over 3}y^{3})$,  although 
in general detailed particle tracking is necessary for determining the 
effective, or `dynamic', aperture of the machine. The same is true conceptually
for linear accelerators comprising a periodic array of quadrupole and/or 
solenoid magnets.
For beams wherein the self-field, {\em i.e.}, space charge, is active, in
general the situation is more complicated.  Transitions in the focusing 
force can lead to dynamics similar to violent relaxation in stellar 
systems, {\em i.e.}, the presence of a hierarchy of collective modes or, for 
large transitions, correlations analogous to plasma turbulence.  Other 
processes, such as beam generation at the source, or bunch compression, 
may impose irregularities in the form of microstructure, {\em i.e.}, density 
correlations, on the beam.  Space-charge forces from these `clumps' may 
influence individual particle orbits in the manner of colored noise.
  
Considerations such as these imply that the microscopic dynamics of 
beams and stellar systems are linked, a point that, until recently, 
seems to have been overlooked in the literature.  The specific mechanism 
that connects these diverse systems is phase mixing.

Section II of this paper describes the models that were considered and the
numerical experiments that were performed. Section III then summarises the
results obtained for unperturbed Hamiltonian systems. Section IV describes
how these results are alterred in the presence of friction and white noise,
which might be expected to mimic discreteness effects in a many-body system.
Section V considers in turn the effects of periodic driving and/or coloured
noise with a finite autocorrelation time, which might be expected to mimic
systematic and/or near-random perturbations that cannot be modeled as
instantaneous kicks. Section VI
summarises the principal conclusions and comments on potential applications.

\section{THE EXPERIMENTS PERFORMED}
\par\noindent
The research described here focused on two- and 
three-degrees-of-freedom Hamiltonian systems of the form
\begin{equation}
H={1\over 2}\left( v_{x}^{2}+v_{y}^{2}+v_{z}^{2} \right) + V(x,y,z).
\end{equation}
Three different classes of potential were considered. The simplest,
which appears to exhibit comparatively generic behaviour\cite{K98},
is a straightforward generalisation of the two-degrees-of-freedom
dihedral potential\cite{D4}, allowing for three free parameters,
$a$, $b$, and $c$:
\begin{displaymath}
V(x,y,z)=-x^{2}-y^{2}-z^{2}+{1\over 4}\left( x^{2}+y^{2}+z^{2} \right)^{2}
\end{displaymath}
\begin{equation}
-{1\over 4}\left(ax^{2}y^{2} + by^{2}z^{2} + cz^{2}x^{2} \right).
\end{equation}
The second is given as a combination of isotropic and anisotropic Plummer 
potentials\cite{KAB}:
\begin{displaymath}
\hskip -.5in
V(x,y,z)=-{1\over (a^{2}+x^{2}+y^{2}+z^{2})^{1/2}}
\end{displaymath}
\begin{equation}
\hskip .5in -{m\over (a^{2}+x^{2}+b^{2}y^{2}+c^{2}z^{2})^{1/2}}.
\end{equation}
For generic choices of parameters $a$, $b$, $c$, and $m$  and `reasonable' 
choices of energy $E$, both these potentials yield phase space hypersurfaces 
admitting a complex coexistence of both regular and chaotic orbits.

The third potential is the sum of two integrable contributions, an 
anisotropic harmonic potential and an isotropic Plummer potential\cite{KS}:
\begin{equation}
V(x,y,z)={1\over 2}(a^{2}x^{2}+b^{2}y^{2}+c^{2}z^{2})-
{M\over \sqrt{r^{2}+{\epsilon}^{2}}},
\label{eq:toy}
\end{equation}
with $r^{2}=x^{2}+y^{2}+z^{2}$ .
This potential is interesting in the sense that, provided that $M$ is not too
large, chaotic orbits with positive Lyapunov exponent only exhibit strong 
exponential sensitivity for the small fraction of time they are at 
comparatively small $r$\cite{KS1}. At larger $r$ the motions are `nearly 
regular.'

Two-degree-of-freedom systems were obtained trivially by considering orbits
with initial conditions $z{\;}{\equiv}{\;}v_{z}{\;}{\equiv}{\;}0$.

Results derived for motion in the unperturbed potentials were contrasted
with motions subjected to two different classes of perturbations:
\par\noindent
1. {\em Periodic driving.} This entailed either allowing for a simple 
sinusoidal driving, resulting in evolution equations of the form
\begin{equation}
{d^{2}r_{i}\over dt^{2}}=-{{\partial}V\over {\partial}r_{i}}+
A_{i}\sin\,{\omega}_{i}t \qquad (i=x,y,z),
\end{equation}
or introducing systematic oscillations in one or more of the constants
in the unperturbed potential, {\em e.g.,} considering a perturbed version
of eq.~(2.3) with 
\begin{equation}
a\to a_{0}(1+\sin\,{\omega}t) \;\;\;{\rm or}\;\;\;
m\to m_{0}(1+\sin\,{\omega}t),
\end{equation}
or
\begin{equation}
V({\bf r},t) \to \left( 1 + A\sin {\omega}t \right) \times V({\bf r}),
\end{equation}
\par\noindent
2. {\em Friction and noise.} This entailed considering Langevin equations
of the form
\begin{equation}
{d^{2}r_{i}\over dt^{2}}=-{{\partial}V\over {\partial}r_{i}}-
{\eta}{dr_{i}\over dt}+F_{i} \qquad (i=x,y,z),
\end{equation}
which incorporate a dynamical friction $-{\eta}dr_{i}/dt$ and a pseudo-random
force $F_{i}$, idealised as Gaussian noise with zero mean. The friction
and noise were typically taken to be related by a Fluctuation-Dissipation
Theorem
\begin{equation}
{\langle}F_{i}(t_{1})F_{j}(t_{2}){\rangle}=K(t_{1}-t_{2}){\delta}_{ij}
\end{equation}
with diffusion constant
\begin{equation}
D=\int_{-\infty}^{\infty}d{\tau}K({\tau})=2{\eta}{\Theta},
\end{equation}
in terms of a `temperature' ${\Theta}{\;}{\sim}{\;}|E|$. 
Three forms of noise were considered, namely:
\par\noindent
${\bullet}$ {\em white noise}, which is delta-correlated in time, with
\begin{equation}
K({\tau})=2{\Theta}{\eta}{\delta}_{D}({\tau}),
\end{equation}
${\bullet}$ {\em coloured noise} sampling an Ornstein-Uhlenbeck process, for 
which the autocorrelation function
\begin{equation}
K({\tau})={\alpha}{\eta}{\Theta}\exp(-{\alpha}|{\tau}|),
\end{equation}
corresponding to an autocorrelation time $t_{c}=1/{\alpha}$, and 
\par\noindent
${\bullet}$ {\em coloured noise} characterised by a more complex 
autocorrelation function
\begin{equation}
K({\tau})={3{\alpha}{\eta}{\Theta}\over 8}\exp(-{\alpha}|{\tau}|)
\left( 1 + {\alpha}|{\tau}|+{{\alpha}^{2}\over 3}{\tau}^{2} \right),
\end{equation}
for which $t_{c}=2/{\alpha}$.
The white noise autocorrelation function (2.10) can be derived from either
(2.11) or (2.12) in a singular limit ${\alpha}\to\infty$. White noise was 
implemented numerically using a standard algorithm\cite{GSH}. Coloured
noise was implemented using an algorithm developed by I. V. Pogorelov
as part of his Ph.~D. dissertation\cite{Pogo}.

Phase mixing was investigated by tracking the evolution of localised ensembles
of $N$ initial conditions of fixed energy $E$, with 
$1600{\;}{\le}{\;}N{\;}{\le}{\;}40000.$
These were typically generated by uniformly sampling a four- or 
six-dimensional phase space hypercube of side ${\Delta}Z$, with 
${\Delta}Z{\;}{\sim}{\;}10^{-3}$ the size of the accessible phase space region.
The initial conditions were integrated using a
standard integration scheme which simultaneously tracked the evolution of a
small initial perturbation, periodically renormalised in the usual 
way\cite{LL1}, so as to extract estimates of the largest (finite time) 
Lyapunov exponent. 

Two diagnostics were used to probe the evolution of each ensemble:
\par\noindent
${\bullet}$ {\em lower order moments} of the form 
\begin{equation}
{\langle}x^{A}y^{B}z^{C}v_{x}^{D}v_{y}^{E}v_{z}^{F}{\rangle} 
\qquad 
A+B+C+D+E+F{\;}{\le}{\;}6,
\end{equation}
computed in the obvious way; and
\par\noindent
${\bullet}$ {\em coarse-grained} two-dimensional {\em distribution functions}
like $f(x,y,t)$ or $f(x,v_{x},t)$, generated by gridding the orbital data at 
fixed instants of time into a $k \times k$ grid with $k$ ranging between
$5$ and $40$. 

The initial divergence of an ensemble was probed by tracking the 
rate at which quantities like the dispersions ${\sigma}_{x}$ or 
${\sigma}_{v_x}$ grow. The approach towards an equilibrium or near-equilibrium 
was probed by tracking (i) the convergence of various
moments towards time-independent `equilibrium' values and (ii) the evolution
of coarse-grained distributions towards a nearly time-independent form. 
For example, it is an empirical fact that, in each potential, chaotic
ensembles typically evolve towards a state in which, modulo finite-$N$
statistics, ${\langle}x{\rangle}=0$ and ${\sigma}_{x}$ assumes
a constant value ${\sigma}_{x0}$. The obvious question, then, is whether
the decay of ${\langle}x(t){\rangle}$ and 
$D{\sigma}_{x}(t){\;}{\equiv}{\;}{\sigma}_{x}(t)-{\sigma}_{x0}$ towards zero 
is power law or exponential, and, for the
case of chaotic ensembles, the degree to which the rates associated with
this decay correlate with the finite time Lyapunov exponents of the orbits.

Convergence of coarse-grained distributions was handled 
analogously\cite{KM}. Motivated by the visual impression that, at least
for chaotic ensembles, the orbits usually approach a nearly
time-independent equilibrium state, the last $q$ snapshots of a sequence of
$p$ snapshots can be used to derive a near-invariant distribution
\begin{equation}
f_{niv}(Z_{i},Z_{j})={1\over q}\sum_{p-q+1}^{p}f(Z_{i},Z_{j},t_{q}),
\end{equation}
and convergence towards $f_{niv}$ can then be probed by computing discretised 
$L^{p}$ norms
\begin{equation}
D^{p}f(Z_{i},Z_{j},t)=\left( 
\sum\sum |f(Z_{i},Z_{j},t)-f_{niv}(Z_{i},Z_{j})|^{p}\over
\sum\sum |f_{niv}({Z_{i},Z_{j})|^{p}} \right)^{1/p}
\end{equation}
for $p=1$ and $2$, with sums extending over the $k\times k$ grid.

\section{UNPERTURBED HAMILTONIAN SYSTEMS}
\subsection{Phase mixing of regular ensembles}

Localised ensembles of initial conditions corresponding to regular orbits
will, when evolved into the future, diverge as a power law in time.
In particular, quantities like the dispersion ${\sigma}_{x}$ 
will typically grow linearly in
time until they become comparable in magnitude to the size of the phase space
region sampled by individual orbits in the ensemble.

This power law divergence is easily understood. Motion in the $x$- (or any 
other) direction is periodic, so that, after a period 
${\tau}_{x,i}=2{\pi}/{\omega}_{x,i}$ each orbit $i$ will return to its initial 
value $x_{0,i}$. In general, however, the periods ${\tau}_{i}$ for nearby 
orbits  will be slightly different, so that when orbit $i$ has returned to 
$x_{0,i}$, a slightly displaced orbit $j$ with different ${\tau}_{j}$ will not 
have returned to its initial $x_{0,j}$. It will instead assume a slightly 
different value $x_{j}$ which, on the average, is slightly further away from 
$x_{0,i}$ than the initial $x_{0,j}$. On the average, the two orbits will have 
moved further apart. (For the special case of a harmonic potential, where all 
the orbits have the same frequencies, there is {\em no} phase mixing.)

Let $t_{cr}{\;}{\sim}{\;}1/{\omega}$ denote a typical periodicity,
${\delta}{\tau}{\;}{\sim}{\;}1/{\delta}{\omega}$ a characteristic `spread' in
periodicities, and $R$ the linear `size' of the configuration space region 
that is accessible to the orbits in the ensemble. The preceding argument then
suggests that ${\sigma}_{x}$ should grow in such a fashion as to satisfy 
${\sigma}_{x}/R{\;}{\sim}{\;}({\delta}{\omega})t.$ Presuming, however, that
${\delta}{\omega}{\;}{\sim}{\;}{\omega}{\Delta}/R$, where ${\Delta}$ is the 
`size' of the region from which the initial conditions were derived,
it follows that
\begin{equation}
{\sigma}_{x}(t){\;}{\sim}{\;}(t/t_{reg})R, \;\; {\rm with} \;\;
t_{reg}({\Delta}){\;}{\sim}{\;}(R/{\Delta})t_{cr}
\end{equation}
or, equivalently,
\begin{equation}
{\sigma}_{x}(t){\;}{\sim}{\;}{\Delta}\,(t/t_{cr}).
\end{equation}

That ${\sigma}_{x}$ grows linearly in time and that, at least approximately, 
${\sigma}_{x}{\;}{\propto}{\;}{\Delta}$ and 
$t_{reg}{\;}{\propto}{\;}1/{\Delta}$ are both illustrated in Fig.~1, the first 
five panels of which track ${\sigma}_{x}(t)$ for ensembles of initial 
conditions evolved in the two-dimensional dihedral potential with $a=1$ and 
energy $E=2.0$. In each case, a cell of $20000$ initial conditions was 
constructed, centered about a fixed phase space point $(x_{0},y_{0},v_{x,0},
v_{y,0})$. 
The ensembles were generated by setting $v_{y}=v_{y,0}$, uniformly sampling 
a configuration space square of size ${\Delta}$, and, for each choice of $x$
and $y$, solving for $v_{x}=v_{x}(E,x,y,v_{y})>0$. 
The five panels correspond to values 
ranging from ${\Delta}=1.0\times 10^{-3}$ to ${\Delta}=5.0\times 10^{-2}$.
The sixth panel exhibits on a log-log plot the best fit growth time $t_{reg}$ 
as a function of ${\Delta}$. The best fit value for the slope is $p=-0.95$, 
which implies a scaling $t_{reg}{\;}{\propto}{\;}{\Delta}^{-0.95}$. A dashed 
line corresponding to slope $p=-1$ is shown for comparison.
Given that, for this potential and energy, $t_{cr}{\;}{\sim}{\;}5-10$ and
$R{\;}{\sim}{\;}1$, the amplitude of $t_{reg}({\Delta})$ is also 
consistent with eq.~(3.1).

When integrated for much longer times, regular ensembles typically exhibit a 
coarse-grained evolution towards an invariant equilibrium distribution. This 
reflects the fact that, over very long times, individual regular orbits will, 
in a time-averaged sense, tend to uniformly populate the {\em KAM} tori to 
which their motions are restricted. 

This convergence can be quantified by tracking the evolution of lower order 
moments for the ensemble, which eventually converge towards equilibrium 
values. 
Examples thereof are exhibited in the first four panels of Fig.~2, which were 
generated from the same ensembles used to create Fig.~1d and e. These 
correspond to collections of regular `box' orbits, with the configuration 
space topology of Lissajous figures. 
The invariant distributions associated with these ensembles are symmetric
with respect to reflection about the $x$- and $y$-axes and, as such, have moments
${\langle}x{\rangle}_{inv}={\langle}y{\rangle}_{inv}=0$. 

It is evident that ${\langle}x(t){\rangle}$ is indeed converging towards zero,
but that the time scale associated with the convergence is very long. 
Also evident is the fact that the time dependence of 
${\langle}x{\rangle}$ is comparatively complex: Strictly speaking, the 
evolution
is neither exponential nor power law in time; and, even at very late times,
$t>4000$, one can see coherent evolutionary effects reflecting the fact that 
the orbits all have well-defined periodicities. 

One final point is also clear.
Just as the size ${\Delta}$ determines the time scale on which ensembles 
disperse initially, so also ${\Delta}$ fixes the time scale on which they 
evolve towards an equilibrium. For the ensemble exhibited in panels Figs.~2a
and b, ${\Delta}$ is $2.5$ times larger than for the ensemble exhibited
in Figs.~2c and d, so that, as illustrated in Fig.~1, the first ensemble 
disperses approximately $(2.5)^{0.95}{\;}{\approx}{\;}2.38$ times as fast. 
However, by replotting in panels e and f the data for Fig.~2a and b over the 
shorter interval $T=(8000/2.38){\;}{\approx}{\;}3360$, it becomes evident
visually that the first ensemble also approaches an equilibrium approximately 
$2.38$ times as fast. 

\subsection{Phase mixing of chaotic ensembles}

Initially localised ensembles of chaotic orbits evolve very differently. 
Considerable variability can be observed but, overall,
chaotic ensembles tend to disperse exponentially at a rate ${\Lambda}
{\;}{\equiv}{\;}t_{cha}^{-1}$ that is comparable to
the largest positive Lyapunov exponent ${\chi}$. Thus, {\em e.g.,}
\begin{equation}
{\sigma}_{x}{\;}{\sim}{\;}{\Delta}\,\exp (t/t_{cha}) \qquad {\rm with}
\qquad t_{cha}{\;}{\sim}{\;}{\chi}^{-1},
\end{equation}
with ${\Delta}$ the `size' of the cell of initial conditions.
This exponential growth implies that chaotic phase mixing is
typically much faster than regular phase mixing.

The physics of chaotic phase mixing is very different from the phase mixing 
of regular ensembles. Aperiodic chaotic
orbits exhibit exponentially sensitive dependence, so
that two nearby initial conditions separated by some phase space distance 
${\delta}Z_{0}$ will, when evolved into the future, typically diverge 
exponentially at a rate that is set by a Lyapunov exponent:
${\delta}Z(t){\;}{\sim}{\;}{\delta}Z_{0}\exp ({\chi}t)$. Applying this 
principle to the ensemble as a whole leads immediately to eq.~(3.3). 

The appropriate ${\chi}$ entering into eq.~(3.3) is a 
{\em finite time Lyapunov exponent}\cite{grass} appropriate for orbits in the 
ensemble for the interval associated with the exponential divergence,
{\em not} the asymptotic Lyapunov exponent, as defined in a $t\to\infty$ limit,
which can be very different. Two different ensembles 
that rapidly spread to occupy the same phase space region might be expected
to have comparable ${\chi}$'s and to mix at comparable rates. However, 
it is possible for different ensembles, each in
principle able to access the same phase space region, to have different
finite time ${\chi}$'s and to mix at very different rates.

If the dispersal of the ensemble be probed in a generic direction, 
$t_{cha}^{-1}$  will typically be comparable to the {\em largest positive 
(finite time) Lyapunov exponent}. In general, a conservative 
nonintegrable potential in $D$ dimensions has $D-1$ unequal positive 
Lyapunov exponents, corresponding to $D-1$ different unstable directions. 
However, the rate at which an ensemble disperses in a specified phase space 
direction will be dominated by the Lyapunov exponent associated 
with the most unstable direction with a nonzero projection in that specified
direction; and that most unstable direction will typically have a nonzero 
projection in a generic,`randomly chosen' direction\cite{liap}.

It is, however, possible for certain directions to mix much more rapidly than 
others. As an extreme example, consider a three-dimensional potential where
motion in (say) the $z$-direction is integrable and decoupled from chaotic
motion in the two orthogonal directions. In this case, quantities like 
${\sigma}_{z}$ and ${\sigma}_{v_z}$ will exhibit behaviour appropriate for 
regular phase mixing whereas quantities like  ${\sigma}_{x}$ or 
${\sigma}_{v_y}$ will in general exhibit the exponental behaviour predicted 
by eq.~(3.3). 

Individual chaotic orbits evolved in a generic potential can be impacted
significantly by such topological obstructions as cantori and/or Arnold 
webs\cite{LL1},
which serve as {\em entropy barriers} to impede -- but not block -- motion 
between different portions of the same connected phase space region. If,
however, the motions of individual members of an ensemble are impeded, the
phase mixing of the ensemble as a whole will in general be alterred. 

As a simple example, consider an ensemble of initial conditions 
corresponding to chaotic orbits which, at early times, are trapped near a 
regular island by cantori. The largest finite time Lyapunov exponents for such 
`sticky'\cite{con} chaotic orbits will in general be substantially
smaller than the exponents for `wildly chaotic' orbits not trapped near
regular islands\cite{KEB}; and this implies that the ensemble will disperse 
more slowly than a `wildly chaotic' ensemble. Even more significant, 
however, is the fact that, until the orbits have escaped 
through the cantori, quantities like ${\sigma}_{x}$ may remain much smaller 
in magnitude than the asymptotic value ${\sigma}_{x,\infty}$ towards which the 
ensemble converges at late times. 

Representative data for a `wildly chaotic' ensemble are exhibited in the left
panels of Fig.~3, the upper curves in which exhibit ${\sigma}_{x}$, 
${\sigma}_{y}$, and ${\sigma}_{z}$ for an ensemble evolved in the
potential (2.4) with $a=1.95$, $b=1.50$, $c=1.05$, and $M=0.0316$. These
curves show considerable short time structure but it is evident that, overall,
the dispersions grow exponentially until they saturate on a scale comparable
to the size of the accessible configuration space region. The exponential 
growth becomes especially apparent if the data are smoothed via box-car
averaging. The lower curves in these panels, 
translated downwards by a distance $\ln {\sigma}=3.0$, illustrate the effect of
averaging the raw data, recorded at intervals ${\delta}t=0.25$, over
$20$ adjacent points. It is apparent visually that the growth rates associated
with ${\sigma}_{x}$ and ${\sigma}_{y}$ are essentially identical. The behaviour
in the $z$-direction is noticeably different, but the overall growth rate 
remains comparable.

Fig.~4 exhibits dispersions for chaotic ensembles evolved in the potential
(2.2) with $a=b=c=1$. In each case, data were recorded at intervals 
${\delta}t=0.5$, and the curves generated by box-car averaging over 5 adjacent 
points. Panels a and b present, respectively, smoothed configuration and 
velocity space dispersions for a `wildly chaotic' ensemble with $E=1.0$, with 
the $y$- and $z$-components translated downwards by, respectively, $\ln 
{\sigma}=5$ and $10$. The dots surrounding the top curves represent unsmoothed 
data. It is evident that, for this ensemble, all six
dispersions grow exponentially at comparable rates. Panel c presents
configuration space dispersions for a `sticky' ensemble with $E=4.0$. 
In this case, one still identifies an initial exponential growth, but that
growth slows for $\ln {\sigma}>-1.0$ or so, and the eventual approach towards
a near-invariant ${\sigma}_{0}$ is accompanied by oscillations not observed
in most `wildly chaotic' ensembles. Panel d exhibits exhibits configuration
space dispersions for a particularly unusual example, in which dispersal in 
the $z$-direction is much slower than in the $x$- and $y$-directions. This
anomalous behaviour reflects the fact that the cell of initial conditions 
was concentrated near the $x-y$ plane, where ${\partial}V/{\partial}z=0$.

Atypical behaviour is especially common for potentials like (2.4), where
chaotic orbits behave in a nearly regular fashion `most of the time.' It is, 
{\em e.g.,} easy to identify ensembles which, at least initially, avoid 
the `highly chaotic' central region and, as such only mix on comparatively 
long time scales. And, similarly, it is easy to construct ensembles which,
because of the (initial) orbital orientations in the central regions, are
considerably more chaotic in some phase space directions than in others.

As a clear example, one can contrast the data for the `wildly chaotic' ensemble
exhibited in the left panels of Fig.~3 with the right hand panel, which
exhibits data for a different, much less
chaotic, ensemble evolved in the same potential with the same energy. For the
left ensemble, the mean value of the largest finite time Lyapunov exponent
at $t=256$ is ${\langle}{\chi}{\rangle}{\;}{\approx}{\;}0.055$; for the right
ensemble, ${\langle}{\chi}{\rangle}{\;}{\approx}{\;}0.022$. It is clear that,
overall, the dispersions grow more rapidly for the wildly chaotic ensemble.
Overlaying plots for the two ensembles also reveals
another significant point: Although the sticky ensemble originally diverges
more slowly, the rates of divergence become comparable when
$\ln {\sigma}_{x}> -2$, the growth rate for the sticky ensemble having 
increased substantially. This increase reflects the fact that, by this time, 
most of the `sticky' orbits have become unstuck.

The initial exponential divergence eventually saturates and, if the
ensembles are probed for longer times, one typically observes a coarse-grained
evolution, again exponential in time, towards an invariant, or near-invariant,
distribution ${\mu}_{niv}$, {\em i.e.,} a statistical (near-)equilibrium. In
particular, reduced distributions $f(Z_{i},Z_{j})$ and various moments 
typically converge exponentially towards (near-)equilibrium values, so that, 
{\em e.g.,}
\begin{equation}
D{\sigma}_{x}(t){\;}{\equiv}{\;}
{\sigma}_{x}(t)-{\sigma}_{x,0}{\;}{\sim}{\;}\exp(-{\lambda}t).
\end{equation}
This near-invariant ${\mu}_{niv}$ appears to correspond to a uniform 
population of those portions of the phase space which are easily accessible to 
the orbits, {\em i.e.,} blocked by neither conservation laws nor significant 
entropy barriers. If, {\em e.g.,} one considers an ensemble of orbits, all 
with the same energy $E_{0}$, evolved on a phase space hypersurface that is 
almost completely chaotic and where topological obstructions are 
unimportant, the distribution towards which the ensemble evolves is 
well approximated by a microcanonical distribution 
${\mu}{\;}{\propto}{\;}{\delta}_{D}(E-E_{0})$.\cite{KSB}

Just as, in many cases, the ensemble initially disperses at roughly the same 
rate in different directions, so also it is often true that the approach 
towards a near-equilibrium proceeds at comparable rates in different 
directions. 
Indeed, at least for `wildly chaotic' ensembles, the rate ${\lambda}$ 
associated with the approach toward a near-equilibrium appears typically to be 
comparatively insensitive to the diagnostics which one chooses to probe. 

As probed by lower
order moments, the rate of convergence is insensitive to choice of phase space
direction, so that, {\em e.g.,} $D\ln {\langle}x{\rangle}$, 
$D\ln {\langle}y{\rangle}$, and $D\ln {\langle}v_{z}{\rangle}$ typically
converge towards zero at comparable rates. The order of the moment is also
unimportant, so that, {\em e.g.,} the convergence rate associated
with $D\ln {\langle}x^{p}{\rangle}$ is relatively insensitive to the choice
of $p$, at least for $p{\;}{\le}{\;}6$. Cross moments also evolve 
similarly, so that, {\em e.g.,} the rates associated with 
$D\ln {\langle}xy{\rangle}$ and $D\ln {\langle}xv_{x}{\rangle}$ are 
comparable. Analogously, the convergence rates associated with different 
reduced distributions like $D^{p}f(x,y)$ and $D^{p}f(x,v_{x})$ are comparable 
to one another and relatively insensitive to the level of coarse-graining, 
{\em i.e.,} the choice of $k$ for the $k\times k$ binning, at least for 
$5{\;}{\le}{\;}k{\;}{\le}{\;}40$. And, perhaps most strikingly, the rates 
associated with $D^{p}f$ are comparable to the rates associated with lower 
order moments. 

Figs.~5 and 6 exhibit the evolution of a variety of different moments and
reduced distributions for two different ensembles evolved in the potential 
(2.2) with $a=b=c=1$. Fig.~5 was generated from a `wildly chaotic'
ensemble with $E=2.0$. Fig.~6 was generated from a `sticky' ensemble with
$E=4.0$. It is evident that all the curves in Fig.~5 exhibit an initial
exponential decrease; and it is also apparent that the slope associated
with different moments and different reduced distributions are
comparable in magnitude. (The approach towards non-zero nearly constant values 
at later times is a reflection of finite $N$ statistics.) By contrast, many
of the curves in Fig.~6 exhibit significant deviations from the linear 
behaviour associated with an exponential decrease; and, even when the data
are well fit by an exponential, the rates can differ appreciably. 

The extent to which the moments for these ensembles
are, or are not, comparable can be gauged visually from
the first two panels of Fig.~7, which record best fit rates ${\lambda}$ for 
$32$ representative quantities. The first 12 quantities, represented by 
diamonds, correspond to
odd-powered moments ${\langle}x{\rangle}$, ${\langle}y{\rangle}$,
${\langle}z{\rangle}$, ${\langle}v_{x}{\rangle}$, ${\langle}v_{y}{\rangle}$,
${\langle}v_{z}{\rangle}$, ${\langle}x^{5}{\rangle}$, 
${\langle}y^{5}{\rangle}$, ${\langle}z^{5}{\rangle}$, 
${\langle}v_{x}^{5}{\rangle}$, ${\langle}v_{y}^{5}{\rangle}$, and
${\langle}v_{z}^{5}{\rangle}$.
The next six, represented by squares, correspond to even-powered moments
${\langle}x^{2}{\rangle}$, ${\langle}y^{2}{\rangle}$, 
${\langle}z^{2}{\rangle}$, ${\langle}v_{x}^{2}{\rangle}$, 
${\langle}v_{y}^{2}{\rangle}$, and ${\langle}v_{z}^{2}{\rangle}$.
Following these are mixed moments
${\langle}xy{\rangle}$, ${\langle}yz{\rangle}$, and
${\langle}zx{\rangle}$, represented by $\times$'s, and
${\langle}xv_{x}{\rangle}$, ${\langle}yv_{y}{\rangle}$, and
${\langle}zv_{z}{\rangle}$, represented by $+$'s. 
The last eight points, represented by triangles, correspond to the quantities
$D^{1}f(v_{x},v_{y})$, $D^{1}f(x,v_{x})$, and $D^{1}f(x,y)$, computed for 
$k=40$ with an $L^{1}$ norm, $D^{1}f(x,y)$ for $k=30,20,$ and $10$, and the
quantities $D^{2}f(x,y)$ and $D^{2}f(v_{x},v_{y})$ computed with an $L^{2}$ 
norm.

For the `wildly chaotic' ensemble used to generate Fig.~7a,
all the moments and most of the reduced distributions converge at similar 
rates, comparable to the mean rate associated with all the moments, which
is represented by the horizontal line. The largest deviations 
are observed for the $L^{2}$ distances $D^{2}f$, where the convergence rates 
are appreciably larger. The `sticky' ensemble in Fig.~7b is more
complex, with the best fit rates exhibiting a much larger variability. 
There are, however, certain regularities. For example, moments like
${\langle}xy{\rangle}$, ${\langle}yz{\rangle}$, and ${\langle}zx{\rangle}$,
which probe two different configuration space directions, converge 
substantially more slowly than  ${\langle}xv_{x}{\rangle}$, 
${\langle}yv_{y}{\rangle}$, and ${\langle}zv_{z}{\rangle}$.
Overall, it would appear that the moments divide empirically into two classes,
corresponding to faster and slower convergence rates. The two vertical lines
represent the mean rates for the moments with rates greater than, and
less than, ${\lambda}=0.042$.

The rates ${\lambda}$ associated with the approach towards equilibrium are
invariable smaller than the typical value of the largest (finite time) 
Lyapunov 
exponent ${\chi}$. However, one often observes strong correlations between 
the values of ${\lambda}$ and ${\chi}$, larger ${\chi}$ correlating with a
more rapid approach towards a near-equilibrium. Significantly, though, this
correlation does {\em not} appear universal.

An example of the connection between ${\lambda}$ and ${\chi}$ is exhibited in
panel (c) of Fig.~7, which exhibits the ratio ${\lambda}/{\chi}$ as a function
of energy $E$ for some typical `wildly chaotic' ensembles evolved in the 
potential (2.2). Diamonds represent values generated by averaging over the 
rates associated with ${\langle}x{\rangle}$, ${\langle}y{\rangle}$, and
${\langle}z{\rangle}$, whereas triangles represent values generated from
$Df(x,y)$, $Df(y,z)$, and $Df(z,x)$. In each case, the error bars reflect
variances associated with the mean values. The ratio ${\lambda}/{\chi}$ seems 
to assume a nearly constant value ${\sim}{\;}0.2$, independent of energy.
However, this ratio exhibits significantly more variability for `sticky 
chaotic' 
ensembles, although it {\em does} appear to remain smaller than unity.

The preceding examples illustrate the fact that, for `wildly chaotic' 
ensembles, one can speak meaningfully of a coarse-grained approach, exponential
in time, towards a near-invariant distribution on a time scale $t_{niv}$ that 
is comparable to, but somewhat longer than, the characteristic Lyapunov time 
${\chi}^{-1}$. For `sticky' ensembles the evolution will be somewhat slower,
and sizeable deviations from a purely exponential evolution may be apparent, 
but again there is a roughly exponential approach towards a near-invariant
distribution. 

In neither case, however, is there a guarantee that this near-invariant 
distribution 
corresponds to a true equilibrium. Because of entropy barriers like cantori or 
an Arnold web, orbits may be impeded temporarily from accessing various phase 
space region which are in principle accessible, so that the ensemble may 
evolve initially towards a distribution that involves a nearly uniform 
population of only part of the accessible phase space hypersurface.
Only over considerably longer time scales is a true equilibrium achieved.

This is illustrated in the first two panels of Fig.~8, which were generated 
from a collection
of 20000 initial conditions with energy $E=6.0$ evolved in the two-dimensional 
dihedral potential with $a=1$. The members of the ensemble, corresponding 
initially to wildly chaotic orbits, rapidly approached a near-invariant
distribution, the form of which can be inferred from Fig.~8a, which exhibits
the spatial coordinates of all the orbits at $t=36.0$. This distribution is 
similar to, but clearly distinct from, the apparent true invariant 
distribution which is only approached on a significantly longer time scale. 
This true invariant distribution, illustrated in Fig.~8b by the spatial
coordinates at $t=720$, is clearly more nearly uniform than the earlier
near-invariant distribution. In particular, the distinctive `x-shaped' 
region of higher concentration has been blurred as orbits diffused away from
the diagonals to occupy regions that were avoided originally.

\section{THE EFFECTS OF FRICTION AND WHITE NOISE}
For sufficiently large amplitudes and/or over sufficiently long time scales,
friction and white noise will dramatically effect the evolution of orbit
ensembles since energy is no longer conserved. For `typical', {\em e.g.,} 
additive, white noise, the energies of individual orbits only change 
significantly on a relaxation time scale $t_{R}={\eta}^{-1}$, as the ensemble 
evolves towards a canonical distribution, 
$f{\;}{\propto}{\;}\exp(-E/{\Theta})$, although various sorts of 
multiplicative noise can dramatically reduce $t_{R}$\cite{LS}. 
The objective here is to assess the effects
of friction and noise over time scales sufficiently short that the energies
of individual orbits are almost conserved. Interest, therefore, will be 
restricted largely to effects proceeding on time scales ${\ll}{\;}{\eta}^{-1}$.

Because random perturbations typically push nearby trajectories apart, 
friction and noise tend generically to increase the efficacy of phase mixing 
for both regular and chaotic ensembles.  

Phase mixing of unperturbed regular orbits arises because nearby orbits
typically oscillate with unequal frequencies. Perturbing the orbits by friction
and white noise gives rise to an additional phase mixing that is very 
different in origin. Friction and noise act to induce a random walk
in phase space, so that an unperturbed orbit and a noisy orbit with the same
initial condition, or two different noisy realisations of the same initial
condition, will diverge, at least initially, as $t^{1/2}$. Consider, {\em 
e.g.,} a harmonic potential, where one can proceed analytically\cite{van}.
Assuming that the natural frequency ${\omega}{\;}{\gg}{\;}{\eta}$,
one can average over oscillatory behaviour proceeding on a time scale
${\omega}^{-1}$ to conclude that, for $t{\;}{\ll}{\;}{\eta}^{-1}$, an ensemble
generated from different noisy realisations of the same initial condition 
will disperse in such a fashion that, {\em e.g.,}
\begin{equation}
{\omega}^{2}{\delta}{x}^{2}=2{\Theta}{\eta}t.
\end{equation}
This is a comparatively weak effect, ${\delta}x$ growing on the same time 
scale as the spread in energies which 
satisfies ${\delta}E^{2}/E_{0}=2{\Theta}{\eta}t$, with $E_{0}$ the initial
energy.

For more generic potentials, where unperturbed orbits oscillate with unequal
frequencies, the evolution can be more complex. Friction and noise can be 
viewed as continually deflecting orbits from one periodic trajectory to 
another, but orbits with different periodicities are susceptible to the linear
phase mixing described in the preceding section. One point, however, {\em is}
clear. For the case of regular orbits, friction and noise can have significant
effect on the {\em bulk} evolution of an ensemble only on the long time scale 
${\eta}^{-1}$ associated with changes in energy. 

Nevertheless, friction and noise {\em can} have noticeable effects over
shorter time scales by `fuzzing out' shorter scale structure. Because regular
orbits are periodic, a plot of ${\sigma}_{x}$ for an unperturbed regular 
ensemble will typically exhibit considerable structure superimposed on an
average linear growth law. The introduction of friction and noise destroys 
this exact periodicity and, consequently, can help suppress this structure.
At a microscopic level, this `fuzzing' is associated with the fact that,
with the introduction of noise, Liouville's Theorem is no longer applicable,
so that it becomes possible for orbits to self-intersect. 
Friction and noise can also accelerate the evolution towards a `well-mixed'
state in which the basic symmetries of the potential are manifest, so that,
{\em e.g.,} for the potentials described in Section II, ${\langle}x{\rangle}=
{\langle}y{\rangle}={\langle}z{\rangle}=0$. However, this is comparatively
slow effect, which only proceeds on the time scale $t_{R}={\eta}^{-1}$ 
associated with significant changes in the energies of individual orbits.
 
As an example, consider Fig.~9, which exhibits $\ln |{\langle}x{\rangle}|$
for an ensemble of regular orbits with initial energy $E=2.0$ evolved in the 
potential (2.2). The first panel,
generated from unperturbed orbits, shows no evidence of evolution towards 
${\langle}x{\rangle}=0$. The remaining three panels, generated from orbits 
evolved with ${\Theta}=E$ and ${\eta}=10^{-7}$, $10^{-6}$, and $10^{-5}$, 
show clear evolutionary effects, but only on a timescale comparable to $t_{R}$.
In each case, the energy dispersion 
is well fit by a linear growth law,
\begin{equation}
{\sigma}_{E}^{2}=A{\Theta}{\eta}E_{0}t,
\end{equation}
with $E_{0}$ the initial energy and $A$ a constant of order unity, so that, 
{\em e.g.,} at time $t=512$, the energy dispersions for the three panels were, 
respectively, ${\sigma}_{E}=0.032$, $0.101$, and $0.318$. It is thus evident 
that, by the 
time $|{\langle}x{\rangle}|$ has begun to evidence a significant systematic 
decrease, the orbital energies have manifested appreciable changes, so
that one can no longer envision an ensemble of orbits restricted, even 
approximately, to a fixed constant-energy hypersurface.

Noise acts to enhance chaotic phase mixing in a very different way. Because
of their exponential sensitivity, perturbed and unperturbed orbits with the
same initial conditions, or two different noisy realisations of the same 
initial condition, will typically diverge at a rate set by an appropriate 
finite time Lyapunov exponent, but with a prefactor that depends on the 
amplitude of the perturbation. In particular, one has\cite{HKM}
\begin{equation}
{\delta}r{\;}{\sim}{\;}\sqrt{{{\Theta}{\eta}\over {\omega}^{3}}}
\exp(t/t_{cha}),
\end{equation}
with $t_{cha}{\;}{\sim}{\;}{\chi}^{-1}$ and ${\tau}{\;}{\sim}{\;}1/{\omega}$ 
a characteristic orbital time scale. 
Noise acts as a `seed' to push apart two orbits with the same
initial condition but, once separated, their separation will grow 
exponentially. The form of the prefactor is easily understood: Noise acts as 
a `random' process, so that its efficacy scales as the square root of 
the amplitude of the perturbation, {\em i.e.,} 
${\propto}\;({\Theta}{\eta})^{1/2}$. 
The remaining dependence on ${\omega}$ follows from dimensional analysis. 

The implication for chaotic mixing is obvious: If the size ${\Delta}$ of the 
initial ensemble is extremely small or if the amplitude of the noise is
comparatively large, noise can have a substantial effect on the rate at 
which quantities like ${\sigma}_{x}$ grow, even for `wildly chaotic' ensembles.
This could, for example, be of practical importance for charged particle
beams, where the aim is to minimise the growth of emittance in a (hopefully)
very narrow beam. If, however, ${\Delta}$ is not that
small, noise will have a comparatively minimal effect on the rate at which
`wildly chaotic' ensembles mix, although it {\em can} prove important by
helping to smooth out short scale structure\cite{Guck}.

For `sticky' ensembles, weak noise can again smooth out short scale structure.
Even more importantly, however, it can make the ensemble less `sticky'.
Indeed, if the noise is 
sufficiently strong, differences between `sticky' and `wildly chaotic' 
behaviour can be completely erased, an otherwise `sticky' ensemble evolving in 
the fashion that appears `wildly chaotic.'

Both these effects are illustrated in Fig.~10, which exhibits ${\sigma}_{x}$ 
for two different chaotic ensembles evolved in the dihedral potential (2.2) 
in the presence of noise of variable amplitude ranging from ${\eta}=0$ to 
${\eta}=10^{-4}$. Inspection of the left hand panels, generated from 
a `wildly chaotic' ensemble with $E=2.0$, shows that the noise has only a 
comparatively minimal effect on the growth of ${\sigma}_{x}$.
By contrast, for the `sticky' ensemble with $E=4.0$ exhibited in the right 
hand panels, noise with amplitude as small as 
${\eta}=10^{-5}$ is sufficient to largely obliterate the irregular structure
that, in the absence of noise, is evident at very early times. For noise
with amplitude as large as ${\eta}^{-4}$, all structure has completely
disappeared. It is evident visually that the top two curves in the Figure,
corresponding to ${\eta}=10^{-4}$, both look `wildly chaotic'.

Weak noise can prove even more important for the convergence of an ensemble 
towards an invariant or near-invariant distribution. Even for wildly chaotic 
ensembles, noise acts to `fuzz out' orbits on short scales, so that the rate 
of approach towards a near-invariant distribution can be accelerated
significantly and, for `sticky' ensembles, this effect can be especially 
dramatic, the noise serving also to suppress structure that is otherwise
manifested in quantities like $\ln |{\langle}x{\rangle}|$ or $Df(x,y)$.
Examples thereof are illustrated in Fig.~11, 
which exhibits $Df(x,y)$ for the same two ensembles used to generate Fig.~10.

It is clear from even a cursory examination of Figs.~10 and 11 that the 
effects of noise on the initial dispersal of an ensemble and its eventual
approach towards a near-invariant distribution must exhibit at most a weak,
roughly logarithmic dependence on ${\eta}$. This is confirmed by Fig.~12, which
plots the rate ${\Lambda}$ with which $f(x,y)$ approaches a near-invariant
distribution as a function of $\log_{10} {\eta}^{-1}$. For ${\eta}<10^{-6.5}$ 
or so, 
{\em i.e.}, $t_{R}>10^{6.5}$, the noise has a comparatively minimal effect.
For larger ${\eta}$ and smaller $t_{R}$, however, there is an obvious
logarithmic dependence. The left-most two points in the Figure, generated for
the shortest $t_{R}$, correspond to slightly higher values of ${\Lambda}$
than would have been expected from a simple extrapolation from the remaining
points. This, however, easily understood, reflecting as it does the fact that,
for such very large values of ${\eta}$, energy nonconservation is important
even on time scales $t{\;}{\ll}{\;}50$.

This logarithmic dependence is reminiscent of the logarithmic dependence on
${\eta}$ observed for noise-enhanced diffusion through cantori\cite{PK}.
Consider, {\em e.g.,} multiple noisy realisations a single initial condition 
corresponding to an orbit trapped near a regular island by cantori. In this
case, one finds typically that, once the orbits have dispersed to fill the
easily accessible regions not blocked by the cantori, they will begin to leak 
through the cantori in a fashion that is well modeled by a Poisson process
with a rate ${\Lambda}$ that scales logarithmically with amplitude.

Finally, it should be observed that, by facilitating diffusion through entropy
barriers, even weak noise can dramatically accelerate the approach towards a
{\em true} invariant distribution. 
This effect is illustrated in the bottom two panels of Fig.~8, which exhibit 
orbital data for the same initial conditions as the top panels but now 
evolved in the presence of noise with ${\Theta}=E$ and ${\eta}=10^{-6}$ and
$10^{-5}$. The data in both these panels were extracted from noisy orbits
at time $t=36.0$. However it is evident visually from a comparison with Fig.~8a
and b that they more closely resemble unperturbed orbits at the much
later time $t=720$ than at $t=36.0$.

\section{THE EFFECTS PERIODIC DRIVING AND COLOURED NOISE}

Provided that the autocorrelation time $t_{c}$ is short compared with the
natural time scale $t_{cr}$ for the orbits in some ensemble, its precise value
is unimportant, and coloured noise has nearly the same effect, both 
qualitatively and quantitatively, as white noise with the same diffusion 
constant $D$. 
However, when $t_{c}$ becomes comparable to $t_{cr}$, the effects of the noise
grow substantially weaker; and, for $t_{c}{\;}{\gg}{\;}t_{cr}$, the
noise has a comparatively minimal effect. Overall, for 
$t_{c}{\;}{\gg}{\;}t_{cr}$ the effects of the noise exhibit a roughly 
logarithmic dependence on $t_{c}$. 

Fig.~13 demonstrates how the evolution of the unperturbed ensemble exhibited 
in Figs.~11b and 12b is changed by incorporating 
Ornstein-Uhlenbeck noise with ${\Theta}=4.0$, ${\eta}=10^{-5}$, and different 
autocorrelation times ranging between $t_{c}=0.01$ and $t_{c}=10.0$. 
For $t_{c}{\;}{\le}{\;}0.33$ the precise value of $t_{c}$ matters little, the 
evolution of $Df$ being virtually identical to what was observed for the case 
of white noise with the same amplitude. Alternatively, for 
$t_{c}{\;}{\ge}{\;}3.33$, the 
effect of the noise is drastically reduced, ${\sigma}_{x}$ and $Df$ 
behaving very nearly as if no noise were present. Fig.~14 exhibits analogues
of Fig.~13, generated now for orbits evolved in the presence of the more
complicated noise with autocorrelation function given by eq.~(2.13). It is
clear visually that the two noises have virtually identical effects.

Figs.~15 a and b quantify this accelerated evolution by tracking the rate 
${\Lambda}$ associated with the approach of $f(x,y)$ towards a 
near-invariant distribution. The data points were obtained for a wildly 
chaotic ensemble with initial energy $E=2.0$ again evolved in the dihedral 
potential, but now allowing for coloured noise with ${\Theta}=1.0$, 
${\eta}=10^{-5}$, and variable $t_{c}$. The top panel was generated for orbits
evolved in the presence of Ornstein-Uhlenbeck noise; the lower for orbits
evolved with noise characterised by the autocorrelation function (2.13).
For $t_{c}{\;}{\ll}{\;}1$, the precise value of $t_{c}$ is largely irrelevant
and ${\Lambda}$ assumes, at least approximately, the value that obtains
for white noise. Alternatively, for $t_{c}{\;}{\gg}{\;}1$ the rate ${\Lambda}$
assumes a near-constant value close to that associated with the evolution of
an unperturbed ensemble. For intermediate values, ${\Lambda}$ is a decreasing
function of the autocorrelation time which scales logarithmically in $t_{c}$.
Also evident is the fact that the two differences noises have virtually
identical effects. That the scaling in $t_{c}$ is again logarithmic and that
the precise form of the noise seems unimportant is again reminiscent of what
has been observed for noise-enhanced diffusion through cantori\cite{PK}.

These effects are easily understood qualitatively, given the expectation that 
noise impacts
individual orbits via a resonant coupling\cite{PK}. Most of the power in an
orbit, either regular or chaotic, is concentrated at frequencies 
${\omega}{\;}{\sim}{\;}2{\pi}/t_{cr}$, with $t_{cr}$ the natural orbital
time scale, and an efficient coupling requires noise with substantial
power at comparable frequencies. Delta-correlated white noise has a flat 
power spectrum and, as such, can couple to more or less anything. However,
the introduction of a finite autocorrelation time suppresses high frequency
power. If, in particular, $t_{c}{\;}{\gg}{\;}t_{cr}$ there is essentially
no power at the natural frequencies associated with the orbits, so that the
noise has a comparatively minimal effect. In this case, the only coupling 
arises via higher order harmonics. 

This interpretation assumes implicitly that the effects of noise are not
associated with a few especially large kicks, but result rather from a
continuous wiggling triggered by `typical' kicks. This seems reasonable
when considering noise as a source to `fuzz' out trajectories within a 
single, easily accessible phase space region; and an analysis of diffusion
through cantori and/or along an Arnold web \cite{PK} 
suggests that, at least for these sorts of entropy barriers, this is also
the case. By continually wiggling the orbits, noise helps them 
`hunt' for holes in cantori. 

This suggests further that the form of the noise should be largely irrelevant
in determining its overall effect on phase space transport. At least within
the class of Gaussian noise, this appears to be true. It
may be possible to fine-tune the noise to have especially large
or small effects but, generically, the details appear to be unimportant.
If, {\em e.g.,} the additive white noise described in the preceding section 
is replaced by multiplicative noise by making ${\eta}$ a function of the phase 
space coordinates, very little is changed. Thus, if the 
constant ${\eta}={\eta}_{0}$ considered there is replaced by a new
${\eta}={\eta}_{0}(v/{\langle}v{\rangle})^{\pm 2}$, with ${\langle}v{\rangle}$
the average speed of the unperturbed orbits, one finds that
the rates at which moments and coarse-grained distributions evolve towards 
a near-invariant distribution remain essentially unchanged. Similarly, the 
form of the colour is comparatively unimportant. For example, noises with 
the autocorrelation functions given by eqs.~(2.11) and (2.12) yield very
similar effects. The presence or absence of an accompanying dynamical friction
also appears largely irrelevant.

That noise acts on orbits via a resonant coupling can be corroborated by
studying how, when subjected to time-dependent perturbations, the unperturbed
energies of the unperturbed orbits are changed: Even though the effects 
described here are not associated with changes in energy {\em per se}, the
degree to which energy {\em is} changed provides a simple probe of the strength
of the coupling. Since coloured noise can be interpreted as a superposition 
of periodic drivings with different frequencies, it is useful to decompose the
noise into its basic constituents by studying individually the effects of 
periodic driving with different frequencies ${\omega}$. The results of such
an investigation is exhibited in the top two panels of Fig.~16. The data in
Fig.~16a, produced for the `sticky' ensemble used to generate Figs.~10 
and 11, was generated by reintegrating the orbits in the presence of a 
perturbation of the form given by eq.~(2.5), with amplitude 
$A_{x}=A_{y}=A_{z}{\;}{\equiv}{\;}A=10^{-2}$ but variable 
${\omega}_{x}={\omega}_{y}={\omega}_{z}{\;}{\equiv}{\;}{\omega}$. 
Each integration proceeded for a total time $t=256$, and the energy dispersion
${\sigma}_{E}$ was recorded at intervals ${\delta}t=0.25$. The quantity
plotted is the maximum value assumed by ${\sigma}_{E}$ during the integration
at times $t>10.0$ (after the decay of any transients).  
Fig.~16b shows analogous data, generated for the same ensemble of initial
conditions but now integrated in a perturbed potential of the form given by
eq.~(2.7).

It is evident in both cases that, for low frequencies, ${\sigma}_{E}$ assumes 
a near-constant value, that there is a sharp increase for intermediate 
frequencies, and that, for higher frequencies, ${\sigma}_{E}$ is again 
comparatively small. For perturbations satisfying eq.~(2,7),
${\sigma}_{E}{\;}{\propto}{\;}1/{\omega}$ for large ${\omega}$; for 
perturbations satisfying eq.~(2.5), ${\sigma}_{E}$ is nearly independent of
${\omega}$ for high frequencies. For perturbations satisfying eq.~(2.7),
the detailed response at intermediate frequency exhibits a comparatively 
complex dependence on ${\omega}$; for those satisfying eq.~(2.5), the
dependence on ${\omega}$ is smoother. For high and low ${\omega}$, the
time dependence of ${\sigma}_{E}(t)$ can exhibit considerable variability; 
but, for intermediate frequencies where the response to the perturbation is 
strong, one observes invariably a response well fit by a growth law 
${\sigma}_{E}^{2}{\;}{\propto}{\;}t$, the same growth law associated with 
(both white and coloured) noise.

The connection with properties of the unperturbed orbits is apparent from
Fig.~16c, which exhibits Fourier spectra generated from an
evolution of the unperturbed ensemble for a time $t=256$ with data recorded
at intervals ${\delta}t=0.05$. The quantities $|x({\omega})|$ (solid curve) 
and $|y({\omega})|$ (dashed) plotted there were generated by computing the 
quantities $|x_{i}({\omega})|^{2}$  and $|y_{i}({\omega})|^{2}$  for each 
orbit individually, and then constructing, {\em e.g.,} a composite
\begin{equation}
|x({\omega})|^{2} = \sum_{i=1}^{N} |x_{i}({\omega})|^{2}.
\end{equation}
It is obvious that these power spectra share certain features with the plots
of ${\sigma}_{E}$. One again sees a flat dependence on ${\omega}$ at low
frequencies, substantial power at intermediate frequencies, and a power law
decrease, ${\propto}{\;}1/{\omega}$, for sufficiently large 
frequencies. The peak frequencies in Fig.~16a coincide with the peak
frequencies associated with the Fourier spectra, which suggests a simple
resonant coupling. The peak frequencies in Fig.~16b occur at somewhat higher
frequencies. This suggests couplings through harmonics may be more important
in this case, which is hardly surprising, given the more complex 
time-dependence entering into (2.7).

Given these results, it is easy to predict the effects of periodic driving
on the evolution of orbit ensembles. Since coloured noise is 
a superposition of drivings with different periodicities, one would expect
the same logarithmic dependence on amplitude; and,
by analogy with the case of noise, one would only expect a significant response
if the frequency ${\omega}$ associated with the driving is comparable to
the natural frequencies of the orbits that are being perturbed. However, one
might anticipate that quantities like the rate ${\Lambda}$ associated with
the approach towards a near-invariant distribution 
would exhibit a more irregular dependence on ${\omega}$ than the relatively
smooth variation of ${\Lambda}(t_{c})$ associated with coloured noise.
All these predictions are in fact correct. 

Fig.~17 contrasts the evolution of an unperturbed ensemble with $E=4.0$ 
integrated in the two-dimensional dihedral potential with the same ensemble 
integrated in the presence of periodic driving of the form (2.7) with 
amplitude $A=10^{-1.5}$ and variable frequencies ranging between 
${\omega}=0.1$ and ${\omega}=30.0$. The left panel exhibits the dispersion 
${\sigma}_{x}$ both for raw data recorded at intervals ${\delta}t=0.05$ (dots) 
and data smoothed via a box-car average over $20$ adjacent points (solid 
lines). The right panel exhibits $Df(x,y)$.
Two points are apparent. First, it is clear that although the driving 
has a significant effect for driving frequencies ${\omega}=3.0$ and 
${\omega}=10.0$, the higher and lower frequencies have a comparatively minimal
effect. This is completely consistent with the fact that, as is evident from
Fig.~15 (b), the strongest coupling between the driving and the unperturbed
orbits is for energies between roughly $2.0$ and $20.0$. The second point 
is that, for ${\omega}=3.0$ and $10.0$, the driving impacts ${\sigma}_{x}$
both by increasing the overall rate of increase and by suppressing the
prominent wiggles which are evident for the unperturbed ensemble and for the
ensembles perturbed with higher and lower frequencies.

\section{DISCUSSION}
This paper has summarised an investigation of phase
mixing in `realistic' Hamiltonian systems which admit both regular and
chaotic orbits, focusing also on the effects of low amplitude perturbations
modeled as periodic driving and/or friction and (in general coloured) noise.

Localised ensembles of regular initial conditions will, when integrated into 
the future, diverge linearly 
in time so that, {\em e.g.,} the dispersion ${\sigma}_{x}{\;}{\sim}{\;}
{\Delta}(t/t_{cr})$, with ${\Delta}$  the `size' of the region sampled by
the initial conditions and $t_{cr}$  a characteristic orbital time scale.
By contrast, initially localised ensembles of `wildly' chaotic orbits diverge
exponentially at a rate which is typically comparable to the magnitude of the
largest finite time Lyapunov exponent for orbits in the ensemble. `Sticky'
chaotic orbits can exhibit more complex behaviour, but again tend to diverge
exponentially. The prefactor associated with the exponential growth of 
quantities like ${\sigma}_{x}$ again exhibits a nearly linear dependence on
${\Delta}$. 

Viewed over longer time scales, chaotic ensembles typically exhibit a
coarse-grained evolution, exponential in time, towards a near-uniform 
population of those easily accessible phase space regions not seriously
obstructed by entropy barriers. For `wildly chaotic' ensembles,
the rate ${\lambda}$ associated with this convergence tends to be 
insensitive to the choice of the diagnostic used to probe the convergence:
different moments and/or coarse-grained distribution functions converge at
comparable rates. For `sticky' ensembles, the convergence can depend more
sensitively on the choice of diagnostic so that, oftentimes, one can 
distinguish between diagnostics characterised by `fast' and `slow' time scales.
This behaviour seems insensitive to the form of the potential. Different
`generic' potentials exhibit the same qualitative behaviour; and similar
behaviour is also observed for `special' potentials like (2.4) where chaotic
orbits behave in a near-regular fashion most of the time. 

If entropy barriers like cantori or an Arnold web are largely irrelevant, this
near-invariant distribution may correspond
(at least approximately) to a true equilibrium, {\em i.e.,} a uniform,
microcanonical population of the accessible phase space regions. If, however,
such barriers are important, this distribution may differ
significantly from a true equilibrium. Only over much longer time scales 
will orbits diffuse through the entropy barriers to approach a true invariant
distribution.

Viewed over sufficiently long time scales, regular ensembles also exhibit a
coarse-grained evolution towards an invariant distribution which, now, 
corresponds to a uniform population of the tori to which their motions are 
restricted. This later time evolution is neither strictly exponential nor 
power law, but one can at least identify a simply scaling in
terms of ${\Delta}$, the `size' of the region sampled by the initial 
conditions.

Weak perturbations modeled as Gaussian white noise can significantly enhance 
the efficiency of chaotic phase mixing, even over time scales sufficiently 
short that energy is almost conserved. In part by relaxing the constraints 
associated with Liouville's Theorem, the noise can help `fuzz out' short scale 
structure; and, by facilitating diffusion through entropy barriers, it can 
dramatically accelerate the approach towards a true invariant distribution. 
As probed by the rate at which an ensembles approach an invariant or 
near-invariant distribution, the efficacy of the noise exhibits a logarithmic 
dependence on amplitude, the same dependence exhibited by noise as a source of 
accelerated diffusion through entropy barriers. 

Coloured noise with an extremely short autocorrelation time $t_{c}$ has 
virtually the same effect as white noise. However, when $t_{c}$ becomes
comparable to the orbital time scale $t_{cr}$, the effects of the 
noise begin to decrease; and, for $t_{c}{\;}{\gg}{\;}t_{cr}$, coloured noise
has only a comparatively minimal effect. For intermediate times, the efficacy 
of coloured noise exhibits a roughly logarithmic dependence on $t_{c}$. The 
detailed form of the noise seems largely irrelevant: all that seems to matter
are ${\eta}$ and $t_{c}$.

The simulations described here also reinforce the 
interpretation\cite{PK} that noise effects chaotic orbits through a resonant
coupling. Noise can be viewed as a superposition of periodic drivings with
different frequencies combined with random phases, so that it is natural to
focus on how a chaotic ensemble is impacted by driving with a single frequency.
However, a plot of the efficacy of drivings with different frequencies, as 
probed by changes in energy induced by the perturbation, shows distinct 
similarities to a plot of the power spectra for the unperturbed orbits. In
particular, the driving has its largest effect for frequencies comparable to,
or somewhat larger than, the frequencies where the spectra have the largest
power. 

Given that noise can be viewed as a superposition of periodic disturbances 
with a variety of different frequencies all combined with random phases, it is
not surprising that periodic driving impacts orbit ensembles in a fashion 
very similar to noise. The only obvious difference is that the dependence
on frequency is more sensitive than the dependence on autocorrelation time
exhibited by coloured noise. 

\acknowledgments
It is a pleasure to thank Court Bohn for his suggestions about, and comments 
on, the manuscript.
Partial financial support was provided by NSF AST-0070809.

\vfill\eject
\pagestyle{empty}
\begin{figure}[t]
\centering
\centerline{
        \epsfxsize=8cm
        \epsffile{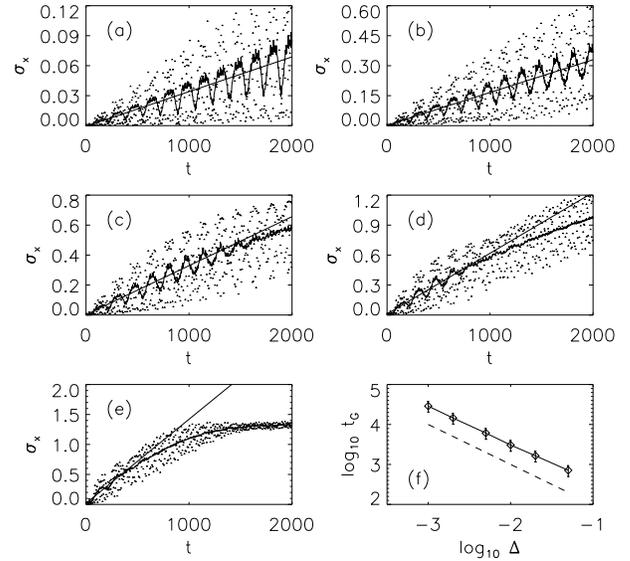}
           }
        \begin{minipage}{10cm}
        \end{minipage}
        \vskip -0.2in\hskip -0.0in
\caption{The dispersion ${\sigma}_{x}$ for ensembles of regular orbits in
the two-dimensional dihedral potential with $a=1$ and energy $E=2.0$, 
generated from $20000$ initial conditions sampling a configuration space 
square of side ${\Delta}$. The dots correspond to data points 
recorded at intervals ${\delta}t=4.0$. The solid curves represent box-car
averages over 10 adjacent points. The straight line exhibits a best fit 
approximation to a linear growth law. (a) ${\Delta}=1\times 10^{-3}$.
(b) ${\Delta}=5\times 10^{-3}$. (c) ${\Delta}=1\times 10^{-2}$.
(d) ${\Delta}=2\times 10^{-2}$. (e) ${\Delta}=5\times 10^{-2}$.
(f) The best fit growth time $t_{G}$ for a linear growth law 
${\sigma}_{x}=t/t_{G}$ as a function of `size' ${\Delta}$. The dashed line
with slope $-1$ corresponds to a proportionality 
$t_{G}{\;}{\propto}{\;}1/{\Delta}$.}
\end{figure}
\vskip -.2in
\begin{figure}[t]
\centering
\centerline{
        \epsfxsize=8cm
        \epsffile{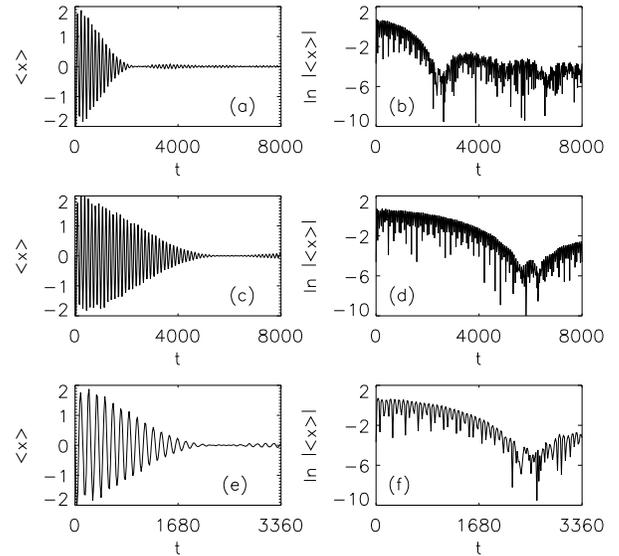}
           }
        \begin{minipage}{10cm}
        \end{minipage}
        \vskip -0.2in\hskip -0.0in
\caption{Convergence of initially localized ensembles of regular orbits
towards an invariant distribution. The ensembles are the same as those used to
generate Fig.~1 d and e, with ${\Delta}=2\times 10^{-2}$ and ${\Delta}=5\times
10^{-2}$. (a) and (b) The mean value ${\langle}x(t){\rangle}$ and 
$\ln |{\langle}x(t){\rangle}|$ for ${\Delta}=5\times 10^{-2}$. (c)
and (d). The same for ${\Delta}=1\times 10^{-2}$. (e) and (f) The same as (a)
and (b) but restricted to a shorter time interval.}
\end{figure}

\begin{figure}[t]
\centering
\centerline{
        \epsfxsize=8cm
        \epsffile{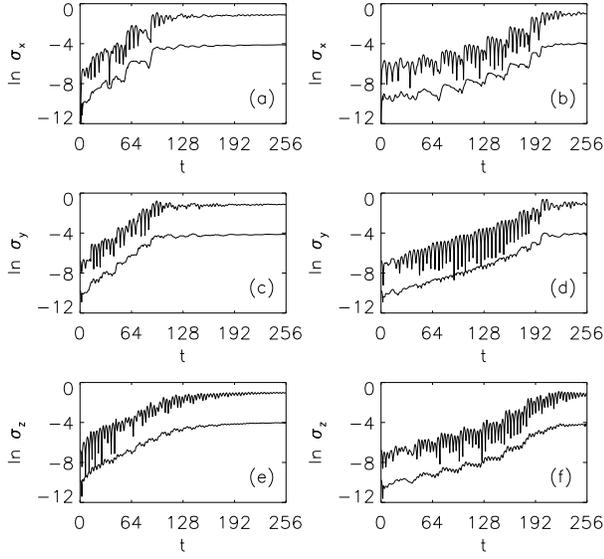}
           }
        \begin{minipage}{10cm}
        \end{minipage}
        \vskip -0.0in\hskip -0.0in
\caption{Smoothed and unsmoothed dispersions ${\sigma}_{x}$ (a and b), 
${\sigma}_{y}$ (c and d), and ${\sigma}_{z}$ (e and f) 
computed for initially localised ensembles of `wildly chaotic' (left
panels) and `sticky' chaotic (right panels) orbits with the same energy
evolved in the ellipsoid plus black hole potential (2.4)}
\end{figure}

\begin{figure}[t]
\centering
\centerline{
        \epsfxsize=8cm
        \epsffile{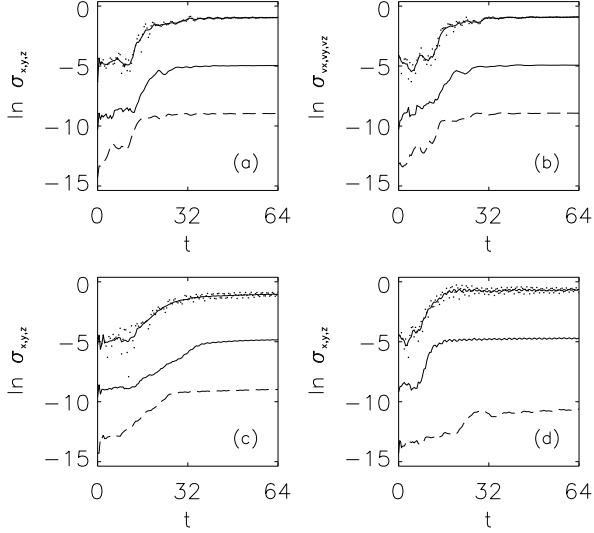}
           }
        \begin{minipage}{10cm}
        \end{minipage}
        \vskip -0.2in\hskip -0.0in
\caption{Smoothed dispersions computed for initially localised ensembles of 
chaotic 
orbits evolved in the potential (2.2) with $a=b=c=1$. (a) ${\sigma}_{x}$ 
(thick solid), ${\sigma}_{y}$ (solid), and ${\sigma}_{z}$ (dashed) for a 
`wildly chaotic' ensemble with $E=1.0$. (b) ${\sigma}_{vx}$, ${\sigma}_{vy}$, 
and ${\sigma}_{vz}$ for the same ensemble. 
(c) ${\sigma}_{x}$, ${\sigma}_{y}$, and ${\sigma}_{z}$ for a `sticky'
chaotic ensemble with $E=4.0$. 
(d) ${\sigma}_{x}$, ${\sigma}_{y}$, and ${\sigma}_{z}$ for a chaotic ensemble 
with $E=4.0$ which is `wildly chaotic' in the $x$- and $y$-directions but 
sticky   in the $z$-direction.}
\end{figure}
\vfill\eject
\begin{figure}[t]
\centering
\centerline{
        \epsfxsize=8cm
        \epsffile{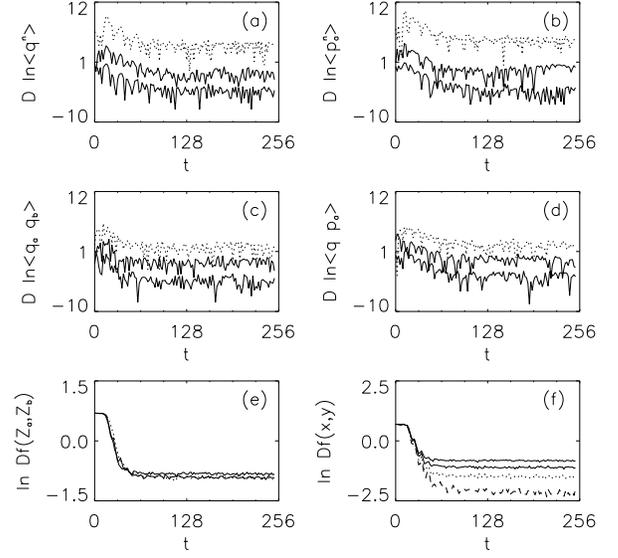}
           }
        \begin{minipage}{10cm}
        \end{minipage}
        \vskip -0.0in\hskip -0.0in
\caption{Convergence towards a near-invariant distribution for an initially
localised ensemble of `wildly chaotic' orbits evolved in the potential (2.2)
with $E=2.0$. (a) Configuration space moments
 $D \ln {\langle}x{\rangle}$ (thick solid),
$D \ln {\langle}y{\rangle}+3$ (thin solid), and 
$D \ln {\langle}z{\rangle}+6$ (dotted). 
(b) Velocity moments $D \ln {\langle}v_{x}{\rangle}$ (thick solid), 
$D \ln {\langle}v_{x}^{2}{\rangle}+3$ (thin solid), and
$D \ln {\langle}v_{x}^{5}{\rangle}+6$ (dotted).
(c) Mixed moments $D \ln {\langle}xy{\rangle}$ (thick solid), 
$D \ln {\langle}yz{\rangle}+3$ (thin solid), and
$D \ln {\langle}xz{\rangle}+6$ (dotted).
(d) $D \ln {\langle}xv_{x}{\rangle}$ (thick solid), 
$D \ln {\langle}yv_{y}{\rangle}$ (thin solid), and
$D \ln {\langle}zv_{z}{\rangle}$ (dotted).
(e) $\ln D^{1}f(x,y)$ (thick solid), $\ln D^{1}f(v_{x},v_{y})$ (thin solid), 
and $\ln D^{1}f(x,v_{x})$ (dotted).
(f) $\ln D^{1}f(x,y)$ computed for different $k \times k$ binnings with (from 
top to bottom) $k=40$, $30$, $20$, and $10$.}
\end{figure}

\begin{figure}[t]
\centering
\centerline{
        \epsfxsize=8cm
        \epsffile{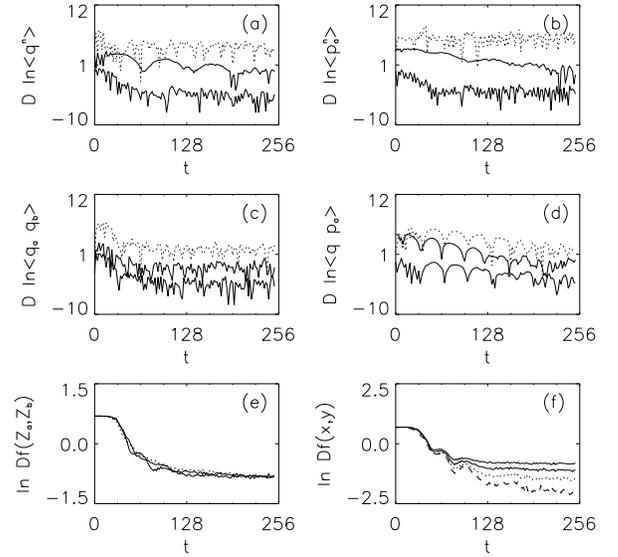}
           }
        \begin{minipage}{10cm}
        \end{minipage}
        \vskip -0.0in\hskip -0.0in
\caption{The same as the preceding Figure, except now generated for a `sticky'
ensemble with $E=4.0$.}
\end{figure}

\begin{figure}[t]
\centering
\centerline{
        \epsfxsize=8cm
        \epsffile{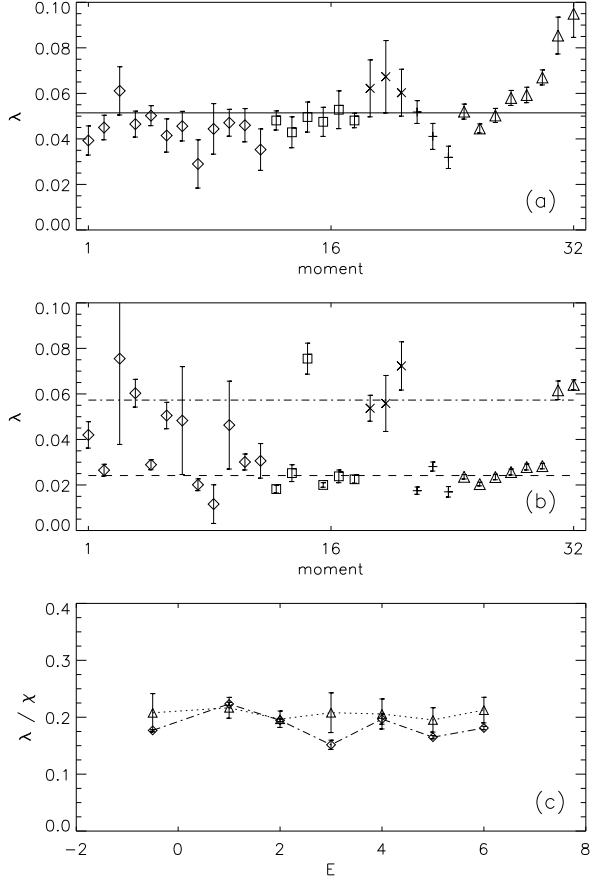}
           }
        \begin{minipage}{10cm}
        \end{minipage}
        \vskip -0.2in\hskip -0.0in
\caption{(a) Representative probes of the rate ${\lambda}$ of convergence 
towards a 
near-invariant distribution for a `wildly chaotic' ensemble with $E=2.0$ 
evolved in the two-dimensional dihedral potential with $a=1$. (b) The same for 
a sticky 
ensemble with $E=4.0$. (c) The ratio ${\lambda}/{\chi}$ for the rates 
${\lambda}$ associated with $D \ln {\langle}q_{a}{\rangle}$ (diamonds) and 
$\ln Df$ (triangles) for
representative `wildly chaotic ensembles' evolved in the dihedral potential 
with different energies.}
\end{figure}
\begin{figure}[t]
\centering
\centerline{
        \epsfxsize=8cm
        \epsffile{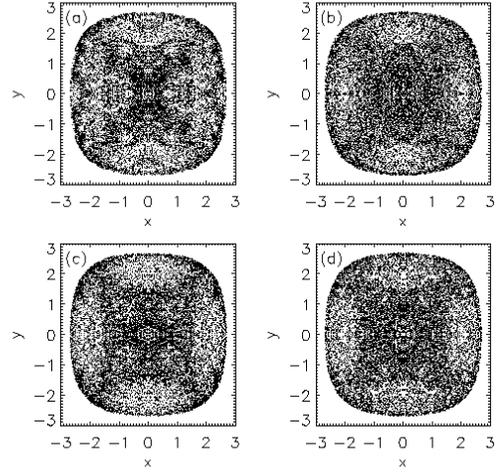}
           }
        \begin{minipage}{10cm}
        \end{minipage}
        \vskip -0.0in\hskip -0.0in
\caption{(a) A collection of 20000 orbits sampling a near-invariant 
distribution with energy $E=6.0$ in the two-dimensional dihedral potential 
with $a=1$ at time $t=36.0$. (b) The same orbits sampled at a much later
time $t=720$. Both ensembles were generated from a localised collection of
initial conditions corresponding to `wildly chaotic' orbits.
(c) The same orbits, now evolved in white noise with ${\Theta}=6.0$ and 
${\eta}=10^{-6}$, again sampled at $t=36.0$. (d). White noise with
${\Theta}=6.0$ and ${\eta}=10^{-5}$, sampled at $t=36.0$.}
\end{figure}
\begin{figure}[t]
\centering
\centerline{
        \epsfxsize=8cm
        \epsffile{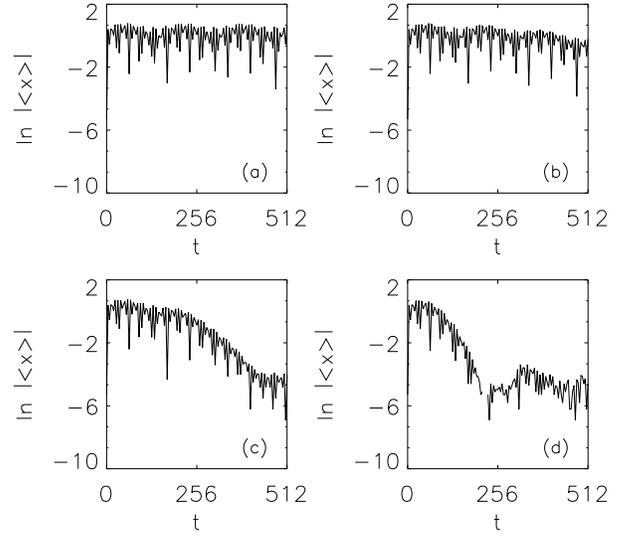}
           }
        \begin{minipage}{10cm}
        \end{minipage}
        \vskip -0.0in\hskip -0.0in
\caption{The quantity $\ln |{\langle}x{\rangle}|$ computed for an initially
localised ensemble of regular orbits with $E=2.0$ evolved in the potential 
(2.2) with $a=b=c=1$. (a) No noise.
(b) ${\eta}=10^{-7}$. (c) ${\eta}=10^{-6}$. (d) ${\eta}=10^{-5}$.}
\end{figure}
\vskip -.2in
\begin{figure}[t]
\centering
\centerline{
        \epsfxsize=8cm
        \epsffile{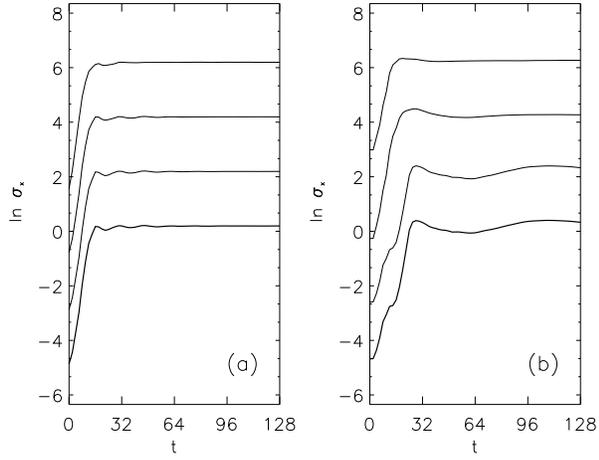}
           }
        \begin{minipage}{10cm}
        \end{minipage}
        \vskip -0.2in\hskip -0.0in
\caption{The dispersion ${\sigma}_{x}$ computed for initially localised 
ensembles of evolved in the two-dimensional dihedral potential with $a=1$ 
in the presence of friction and white noise with ${\Theta}=E$ and (from bottom
to top) variable ${\eta}=0$, $10^{-7}$, $10^{-5}$, and $10^{-4}$. The three
upper curves were shifted upwards by amounts $2$, $4$, and $6$. (a) A `wildly 
chaotic' ensemble with $E=2.0$. (b) A `sticky' ensemble with $E=4.0$.}
\end{figure}

\begin{figure}[t]
\centering
\centerline{
        \epsfxsize=8cm
        \epsffile{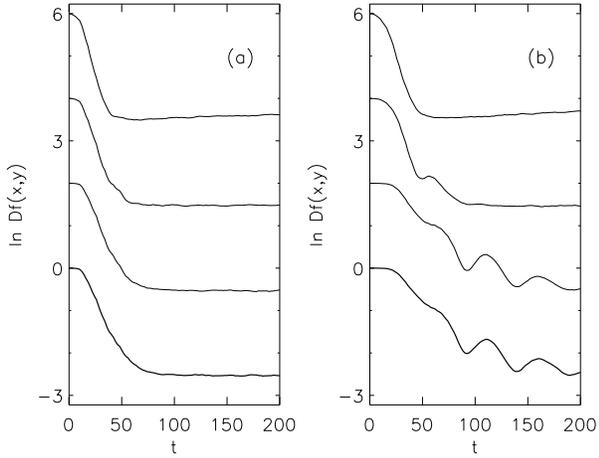}
           }
        \begin{minipage}{10cm}
        \end{minipage}
        \vskip -0.2in\hskip -0.0in
\caption{The quantity $Df(x,y)$ computed  for the `wildly chaotic' (a) and
`sticky' (b) ensembles used to generate
the preceding Figure, again staggered by translations of $2$, $4$, and $6$.}
\end{figure}
\vskip -.5in
\begin{figure}[t]
\centering
\centerline{
        \epsfxsize=8cm
        \epsffile{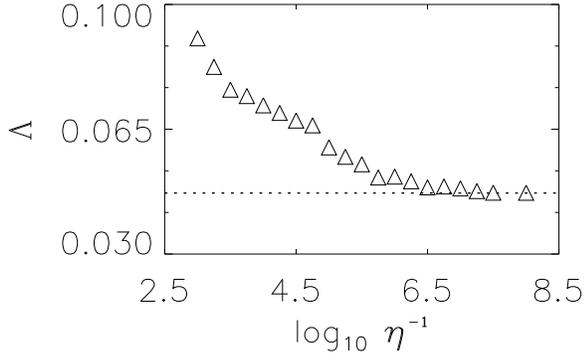}
           }
        \begin{minipage}{10cm}
        \end{minipage}
        \vskip -0.2in\hskip -0.0in
\caption{${\Lambda}$, the rate at which $f(x,y)$ approaches a near-invariant 
distribution, computed as a function of ${\eta}$ for the `wildly chaotic' 
ensemble used to generate Fig. 11. The dashed line corresponds to the zero
noise limit.}
\end{figure}

\begin{figure}[t]
\centering
\centerline{
        \epsfxsize=8cm
        \epsffile{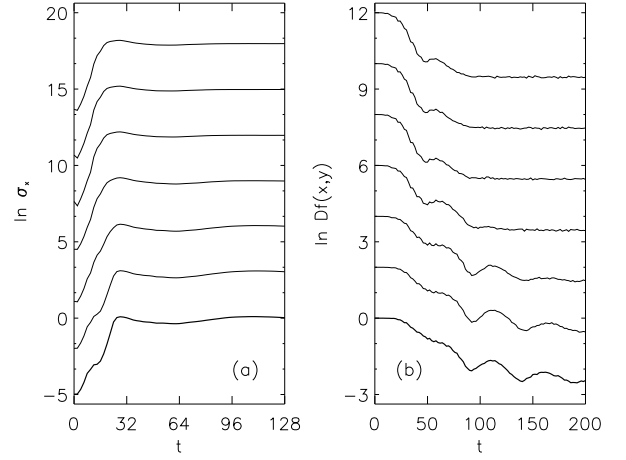}
           }
        \begin{minipage}{10cm}
        \end{minipage}
        \vskip -0.2in\hskip -0.0in
\caption{(a) The dispersion ${\sigma}_{x}$ computed for the ensemble used
to generate Figs.~11 b and 12 b, now integrated in the presence of
Ornstein-Uhlenbeck noise with ${\Theta}=4.0$, ${\eta}=10^{-5}$ and 
(from top to bottom) ${\alpha}=100.0$, ${\alpha}=10.0$, ${\alpha}=3.33$, 
${\alpha}=1.0$, ${\alpha}=0.33$, and ${\alpha}=0.1$,
corresponding to autocorrelation times $t_{c}=0.01$, $0.1$, $0.33$, $1.0$,
$3.33$, and $10.0$. (b) The quantity $\ln |{\langle}x{\rangle}|$ computed
for the same ensembles.}
\end{figure}

\begin{figure}[t]
\centering
\centerline{
        \epsfxsize=8cm
        \epsffile{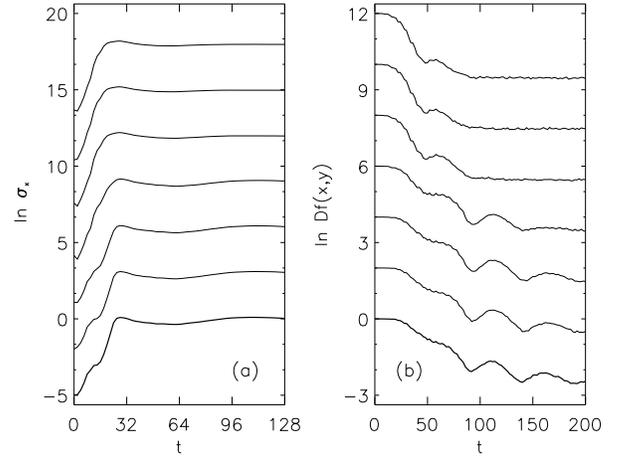}
           }
        \begin{minipage}{10cm}
        \end{minipage}
        \vskip -0.2in\hskip -0.0in
\caption{The analogue of Fig.~13, now computed for orbits in the presence
of coloured noise characterised by the more complex autocorrelation function
(2.12).}
\end{figure}

\begin{figure}[t]
\centering
\centerline{
        \epsfxsize=8cm
        \epsffile{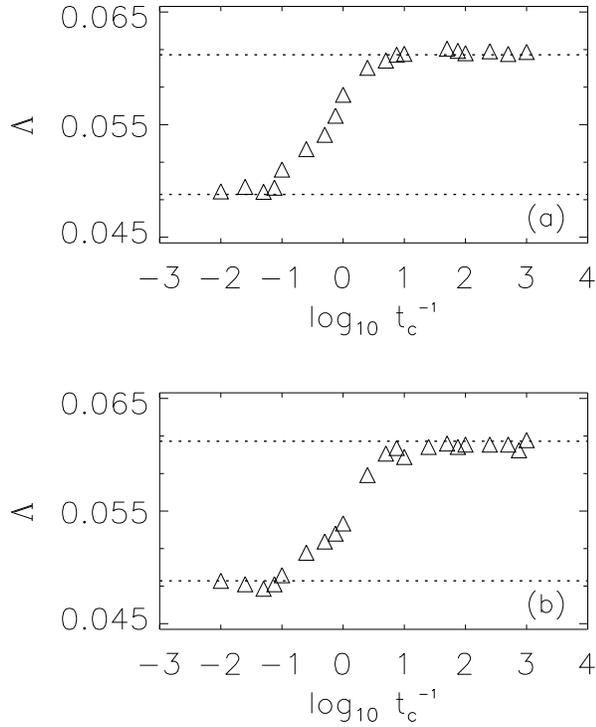}
           }
        \begin{minipage}{10cm}
        \end{minipage}
        \vskip -0.2in\hskip -0.0in
\caption{(a) ${\Lambda}$, the rate at which $f(x,y)$ approaches a 
near-invariant distribution, computed as a function of $t_{c}$ for a `wildly 
chaotic' ensemble with $E=2.0$ evolved in the presence of Ornstein-Uhlenbeck 
noise with ${\Theta}=1.0$ and ${\eta}=10^{-5}$. The lower dashed line 
corresponds to evolution in the absence of noise. The upper line corresponds 
to the $t_{c}=0$ white noise limit. (b) The analogue of (a), now computed for
noise with the autocorrelation function (2.12).}
\end{figure}
\vskip -.4in
\vfill\eject
\begin{figure}[t]
\centering
\centerline{
        \epsfxsize=8cm
        \epsffile{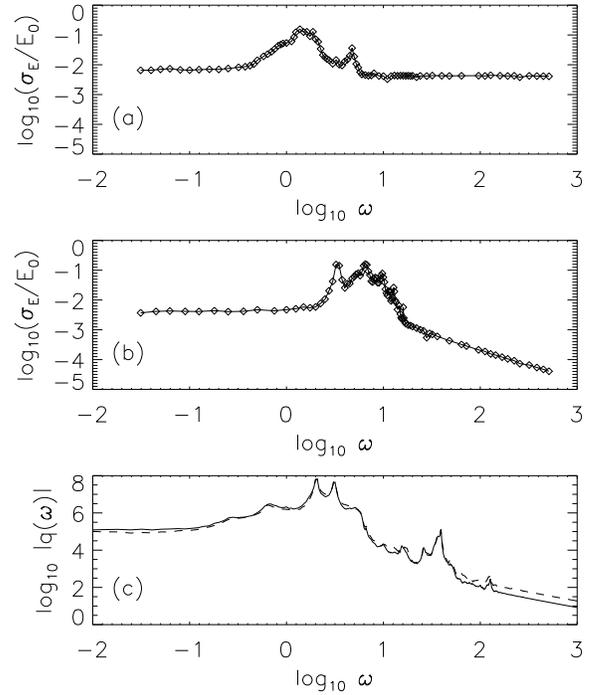}
           }
        \begin{minipage}{10cm}
        \end{minipage}
        \vskip -0.2in\hskip -0.0in
\caption{(a) The maximum value of ${\sigma}_{E}$ for the ensemble
used to generate Figs. 10 and 11b, now evolved for a time $t=256$ in the
presence of periodic driving of the form given by eq.~(2.7) with fixed 
amplitude $A=10^{-2}$ but different frequencies ${\omega}$, expressed in 
units of the original energy $E_{0}=4.0$. (b) The same for driving of the
form given by eq.~(2.5), again with $A=10^{_2}$.
(c) The composite power spectra, $|x({\omega})|$ (solid curve) and 
$|y({\omega})|$ (dashed) computed for the unperturbed ensemble.}
\end{figure}
\vskip -.2in
\begin{figure}[t]
\centering
\centerline{
        \epsfxsize=8cm
        \epsffile{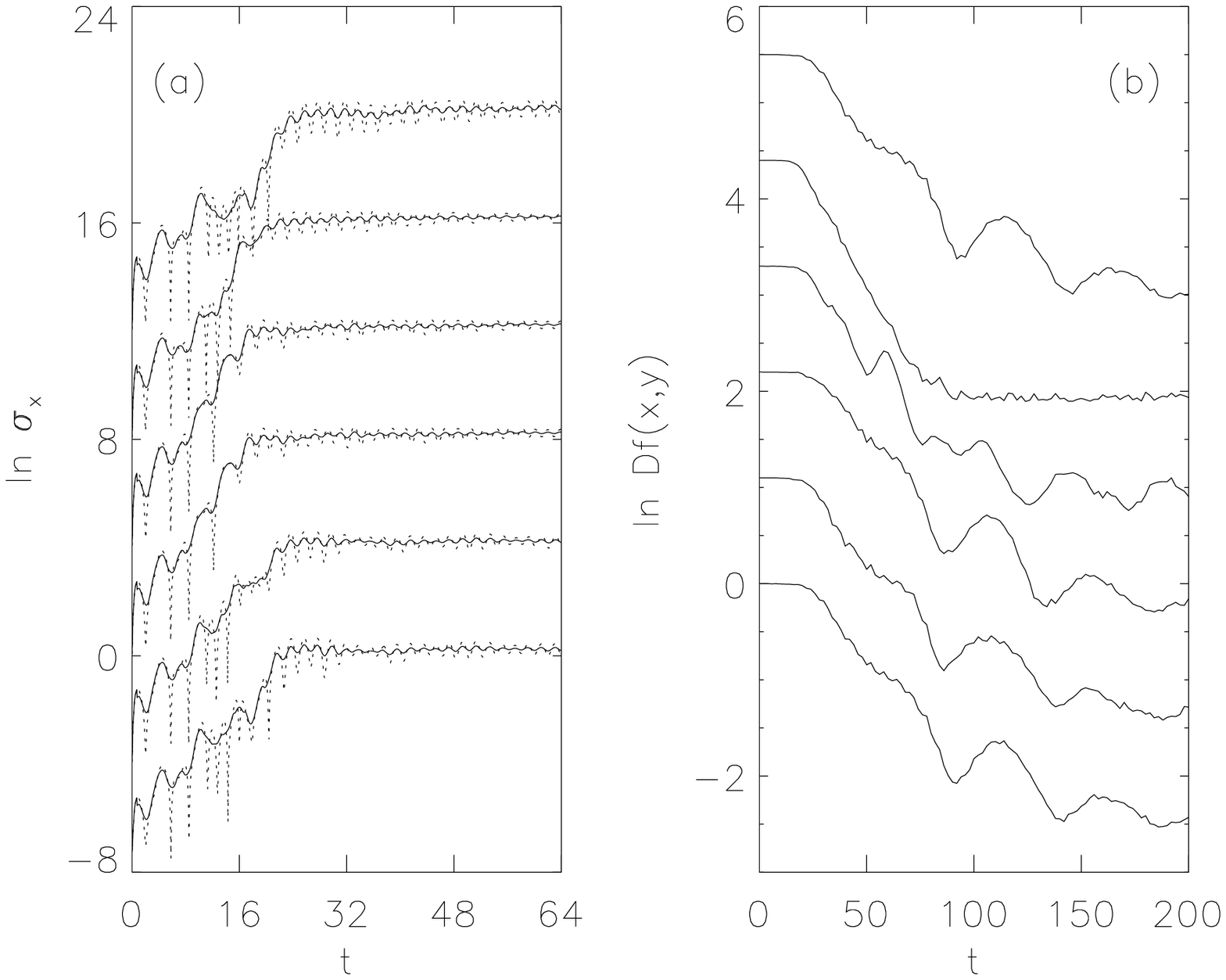}
           }
        \begin{minipage}{10cm}
        \end{minipage}
        \vskip -0.3in\hskip -0.0in
\caption{(a) The dispersion ${\sigma}_{x}$ computed for an ensemble evolved
in the two-dimensional dihedral potential with $a=1$ in the absence of all
perturbations (lower curve) and in the presence of periodic driving of the
form (2.7) with $A=10^{-1.5}$ and (from bottom to top) ${\omega}=0.1$,
${\omega}=1.0$, ${\omega}=3.0$, ${\omega}=10.0$, and ${\omega}=30.0$. The 
upper curves were shifted upwards by amounts $4$, $8$, $12$, $16$, and
$20$. (b) The quantity $\ln |{\langle}x{\rangle}|$ for the same ensembles, the 
upper curves again staggered.
}
\end{figure}


\begin{references}
\bibitem{pos}If one works in terms of the Eisenhart metric, where the 
connection to physical phase space is most direct, the curvature can be
negative only if the second derivative of the potential becomes negative.
\end{references}
\end{document}